# Reduced Dimensionality (4,3)D-hnCOCANH Experiment: An Efficient Backbone Assignment tool for NMR studies of Proteins

Dinesh Kumar*

Centre of Biomedical Magnetic Resonance, SGPGIMS Campus, Raibareli Road, Lucknow-226014, India;

*****Email: dineshcbmr@gmail.com

**ABBREVIATIONS:** NMR, Nuclear Magnetic Resonance; HSQC, Heteronuclear Single Quantum Correlation; CARA, Computer Aided Resonance Assignment; RD: Reduced dimensionality.

**Abstract:**

Sequence specific resonance assignment and secondary structure determination of proteins form the basis for variety of structural and functional proteomics studies by NMR. In this context, an efficient standalone method for rapid assignment of backbone ($^1$H, $^{15}$N, $^{13}$C$^\alpha$ and $^{13}$C') resonances and secondary structure determination of proteins has been presented here. Compared to currently available strategies used for the purpose, the method employs only a single reduced dimensionality (RD) experiment -(4,3)D-hnCOCANH and exploits the linear combinations of backbone ($^{13}$C$^\alpha$ and $^{13}$C') chemical shifts to achieve a dispersion relatively better compared to those of individual chemical shifts (see the text) for efficient and rapid data analysis. Further, the experiment leads to the spectrum with direct distinction of self (intra-residue) and sequential (inter-residue) carbon correlation peaks; these appear opposite in signs and therefore can easily be discriminated without using an additional complementary experiment. On top of all this, the main strength of the method is its efficient and robust assignment strategy based on: (a) multiple, but unidirectional sequential $^{15}$N and $^{13}$C correlations (*i*-->*i*-1) and (b) the distinctive peak patterns of self and sequential peaks in different [($F_1$($^{13}$C)-$F_3$($^1$H) and ($F_2$($^{15}$N)-$F_3$($^1$H)] planes of the 3D spectrum which enable ready identification of certain specific triplet sequences (see the text) and thus serve as check points for mapping the stretches of sequentially connected HSQC cross peaks on to the primary sequence for assigning the resonances sequence specifically. Overall, the standalone method presented here will be an important backbone assignment tool for structural and functional proteomics, protein folding, and drug discovery research programs by NMR.

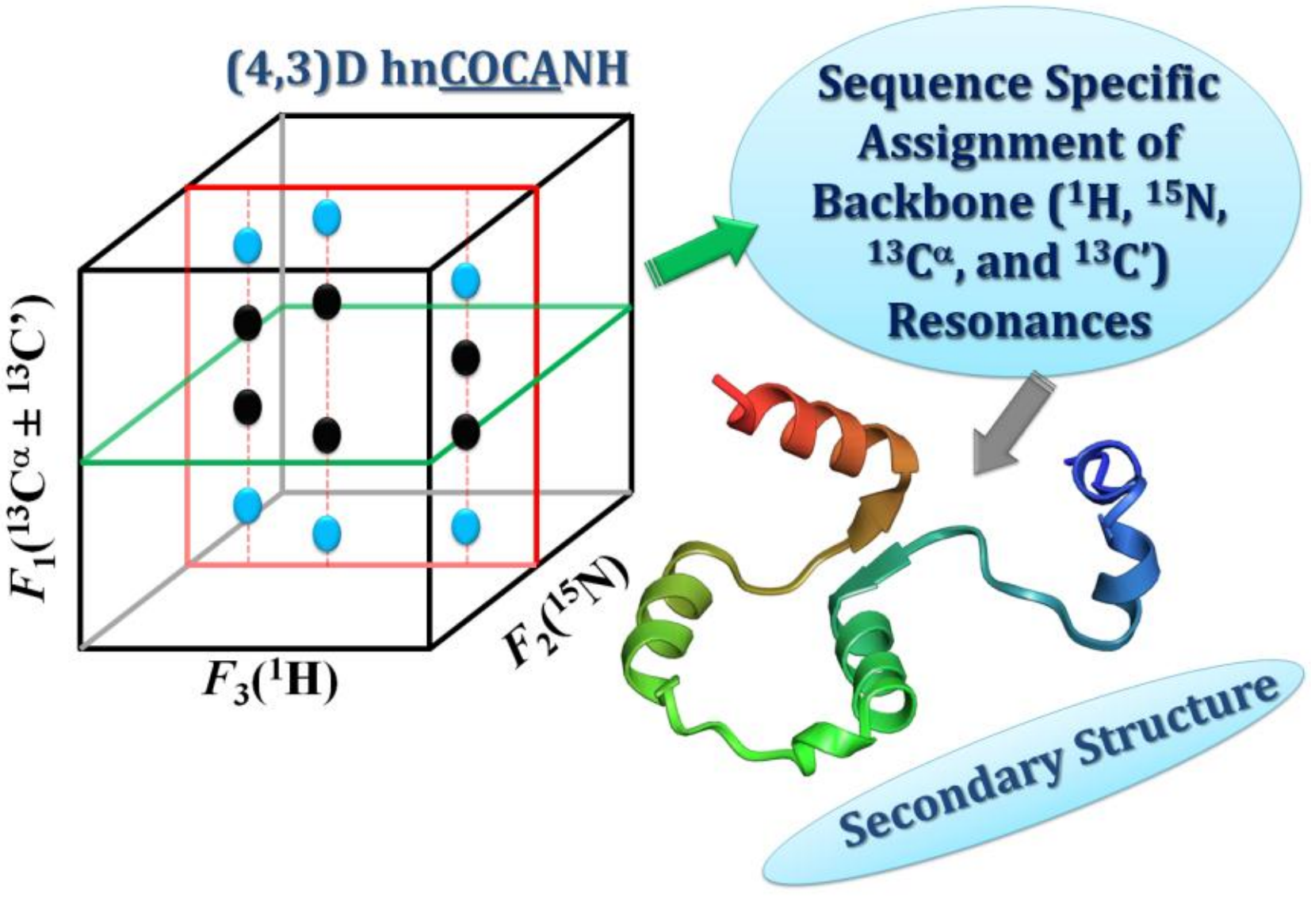

## Introduction:

Backbone resonance assignment and secondary structure determination of proteins form the basis for variety of structural and functional proteomics studies by NMR e.g. (a) three dimensional structure determination of proteins, (b) studying their interaction with various binding partners (like nucleic acids, peptides/proteins, carbohydrates, metal ions, drugs, etc) and (c) studying the effect of mutations on protein structure (which in turn reflects in its bioactivity) and to ascertain the role of specific residues in a particular function [1-3]. NMR further provides a powerful approach for (a) studying protein folding/unfolding pathways and mechanisms especially in the context to understand relationship between a protein's sequence and its tertiary structure [4-6] and (b) characterization of structural and dynamics features of intrinsically unstructured proteins (IUPs; which are not amenable to X-ray crystallographic approaches at all) [4,7]. NMR is also playing a strategic role in drug discovery research programs like (a) high-throughput screening of potential drug candidates based on the amide chemical shift changes in 2D $^{1}$H-$^{15}$N HSQC spectrum of target protein [8], (b) structure based drug discovery (SBDD) programs through the determination of high quality structures of proteins and their complexes [9], (c) fragment based drug discovery (FBDD) programs through identification of the binding site on the target protein and the binding fragments [10,11]. The very first and key requirement in all above studies by NMR is the backbone resonance assignment.

Several NMR experiments and based on strategies have been proposed in the past for backbone resonance assignments in $^{13}$C/$^{15}$N labeled proteins [12-14]. However, these strategies are highly time-consuming, laborious and low-throughput either in terms of data acquisition or data analysis. The major drawbacks are following:

- Most of the currently available strategies involve recording of more than one triple resonance 3D NMR experiments [12] for the assignment of backbone ($^{1}$H$^{N}$, $^{15}$N, $^{13}$C$^{\alpha}$ and $^{13}$C') resonances. However, the use of multiple spectra elaborates the analysis and sometimes may lead to complications due to inter-spectral variations of chemical shifts. The situation –common in proteins which are unstable in solution- further impairs the performance of automated assignment algorithms. Therefore keeping the total number of spectra small always remains highly desirable.

- Most of the established strategies of backbone resonance assignment involve complimentary sets of 3D NMR experiments [14] e.g. HNCA/ HN(CO)CA [15,16], HN(CA)CO/HNCO [17], and CBCANH [18]/CBCACONH [19], HNN/HN(C)N [20]) where the first experiment of each pair needs a complementary one for the discrimination of self and sequential correlations. Further, at least two pairs of such complementary experiments are typically required to establish the complete backbone ($^{1}H^{N}$, $^{15}N$, $^{13}C^{\alpha}$ and $^{13}C'$) assignment. The additional experiments not only increase the demand for NMR instrument time (typically ranging from few days to weeks), but also imposes a condition of long time stability on protein samples. However, there are proteins which are unstable in solution and tend to precipitate in matter of days, thereby reducing the time available to record NMR data. The situation is quite common while studying mammalian as well as intrinsically unstructured proteins where the inherent conformational flexibility renders decreased stability due to destructive intermolecular interactions. Protein stability is also a serious problem during protein folding/unfolding studies by NMR under different concentrations/strengths of a folding/unfolding reagent.

- The routine assignment strategies based on standard 3D NMR experiments (like HNCA, HN(CA)CO or CBCANH) involve repeated scanning through the $^{15}N$ planes of the 3D spectra to search for the matching peaks along the carbon dimension. These strategies work well in case of folded proteins which exhibit good $^{13}C$ chemical shift dispersion. However, in case of proteins with only limited $^{13}C$ chemical shift dispersion and high degree of amide shift degeneracy (e.g. unfolded or intrinsically unstructured proteins and proteins containing repetitions of identical amino acid stretches) the sequential assignments become a challenging task and sometimes may fail.

These are the drawbacks which limits the utility of NMR for high-throughput structural and functional proteomics studies and for modern day drug discovery programs. Thus there is very high demand to develop novel NMR methods and protocols to overcome these limitations. The following key objectives have been identified in this scenario: (i) reducing the number of NMR experiments to obtain the required information, (ii) to increase the speed of data collection, and (iii) to design simple and efficient data analysis strategies.

In the above context, two novel pulse sequences –hncoCANH and hnCOcaNH [21]– were reported previously by our group for rapid and unambiguous assignment of backbone ($^{1}H^{N}$, $^{15}N$, $^{13}C^{\alpha}$ and $^{13}C'$) resonances of proteins. These experiments lead to spectra equivalent to HNCA and HN(CA)CO spectra, respectively, but with direct distinction of inter- and intra-residue correlation peaks (as these appear opposite in signs in the resulted spectra) and thus these do not require any additional complementary experiment. The other advantageous feature of these spectra is the presence of special patterns of self and sequential peaks that are obtained around glycines and prolines. These special patterns enable direct identification of certain specific triplets of residues and therefore provide internal checks for the sequential assignment procedure and explicit side chains assignment becomes less crucial for unambiguous backbone assignment. Despite of having useful spectral features and efficient assignment strategy, the complete assignment of backbone ($^{1}H$, $^{15}N$, $^{13}C^{\alpha}$ and $^{13}C'$) resonances requires recording of both the experiments, where they also complement each other for resolving the ambiguities arising because of degenerate $^{13}C^{\alpha}$/$^{13}C'$ chemical shifts. However, the use of two spectra may even elaborate the analysis and lead to complications due to inter-spectral variations of chemical shifts. In this backdrop, a single experiment (4,3)D-hnCOCANH has been presented here for efficient and rapid assignment of backbone resonances. The method is basically an improvisation of our previous assignment method based on hnCOcaNH and hncoCANH experiments [21] and makes use of the simple reduced dimensionality (RD) [23,24] concept, simultaneously (i) to retain all the beneficial spectral features of hnCOcaNH/hncoCANH experiments and (ii) to achieve higher dispersion along the $F_1$($^{13}$C) axis through linear combination of $^{13}C^{\alpha}$ and $^{13}C'$ chemical shifts **(Fig. 1).**

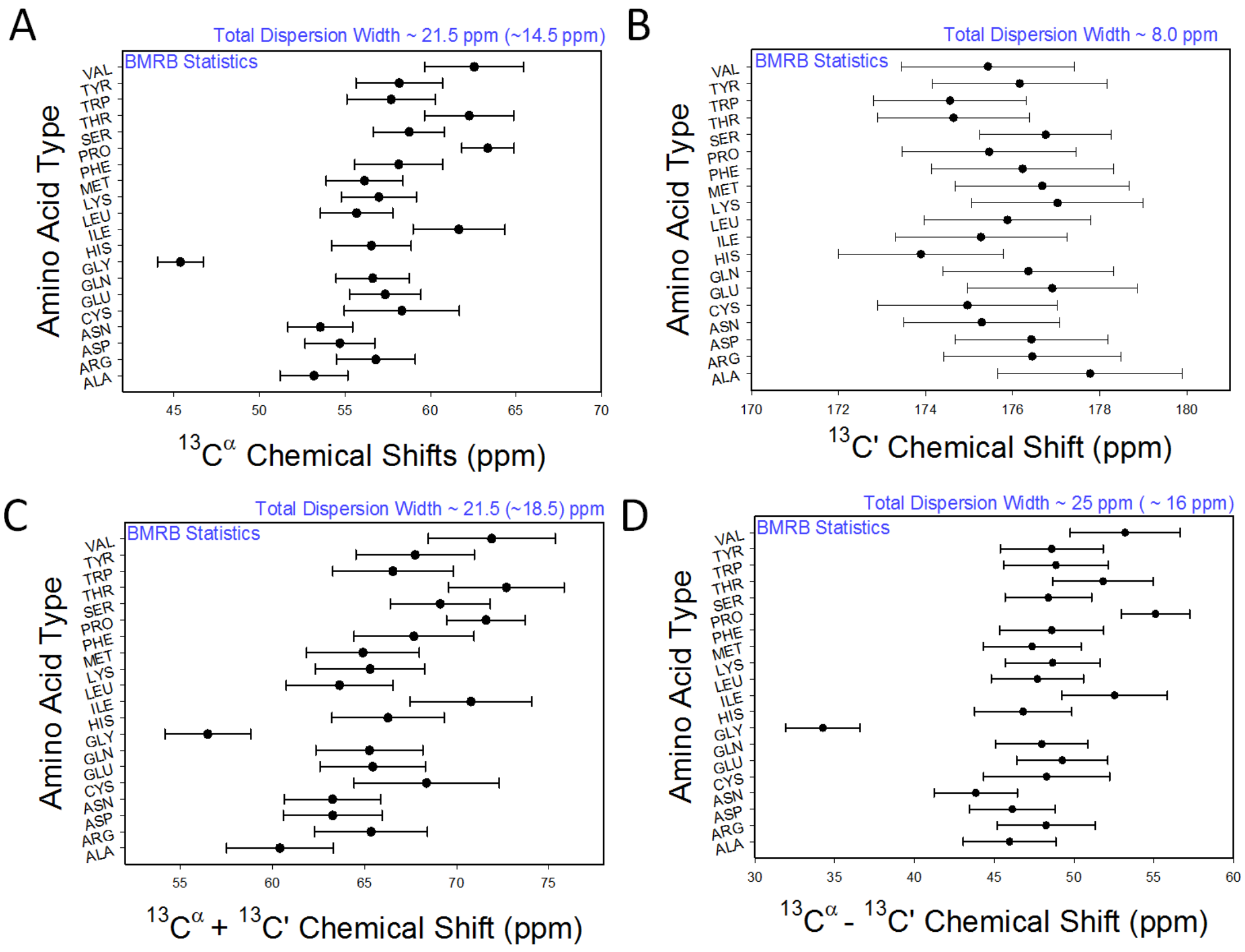

**Figure 1:** Comparison between dispersion of individual $^{13}C^{\alpha}/^{13}C'$ chemical shifts and that of linear combination of these chemical shifts for 20 common amino acids. In **(A)** and **(B)** the average chemical shifts of $^{13}C^{\alpha}$ and $^{13}C'$ are plotted against the residue types. The average chemical shifts have been taken from the BMRB statistical table containing values calculated from the full BMRB database (http://www.bmrb.wisc.edu/ref info/statful .htm#1) [22]. This includes paramagnetic proteins, proteins with aromatic prosthetic groups, and entries where chemical shifts are reported relative to uncommon chemical shift references. Standard deviations in these values are plotted as error bars. In **(C)** and **(D)**, linear combination of $^{13}C^{\alpha}$ and $^{13}C'$ average chemical shifts for all the amino acid types calculated using **Eqns 3 and 4** (given in the main paper) are plotted against the residue types. The deviations shown in **1C** and **1D** were calculated as: $\Delta\delta(^{13}C) = [\Delta\delta(^{13}C^{a})^{2} - \Delta\delta(^{13}C')^{2}]^{1/2}$ where $\Delta\delta(^{13}C^{a})$ and $\Delta\delta(^{13}C')$ are the standard deviations for the individual chemical shifts (plotted in **1A** and **1B**) taken from the BMRB statistical table. The total chemical shift dispersion possibly achieved in each case has been shown at the top of each plot (the values within bracket indicate the dispersion calculated excluding glycines). Comparison clearly shows that the linear combination of $^{13}C^{\alpha}$ and $^{13}C'$ chemical shifts can provide better dispersion compared to the individual chemical shifts.

Following sequence-specific backbone assignment, the $^{13}C^{\alpha}$ and $^{13}C'$ chemical shifts are used to obtain the secondary structural information about the protein using the routine chemical shift index (CSI) method [25-27]. Overall, this whole exercise including data collection and data analysis can be accomplished in 1-3 days depending upon the size, concentration, and the conformational state (folded/unfolded) of the protein in solution. Therefore, the method has wide range of useful applications in NMR investigations of both folded and functional unfolded proteins e.g. (a) high throughput structural proteomics (especially when the protein structure is determined using algorithms based on backbone chemical shift data alone like CS23D [28], CS-Rosetta [29,30] etc), (b) protein folding/unfolding studies where the secondary structure/structural- propensities of the polypeptide chain are characterized at different concentrations/strengths of the unfolding/folding reagent like temperature, pressure, pH, chemical denaturants etc and (c) functional genomics and drug discovery programs where the method can be employed to re-establish the lost assignment as well as to monitor the changes in protein conformation upon ligand/drug binding or upon a mutation.

## Materials and Methods:

The proposed reduced dimensionality pulse sequence -(4,3)D-hnCOCANH- **(Fig. 2)** has successfully been tested and demonstrated on bovine apo Calbindin-d9k protein (75 amino acids, MW = 8.5 kDa). The $^{13}C/^{15}N$ labeled protein sample (1 mM in concentration, 50 mM Ammonium Acetate, pH 6.0, 10% $D_2O$, in high quality NMR tube sealed under inert atmosphere) has been purchased from CIRMMP (Interuniversity Consortium for Magnetic Resonance, Itlay: *http://www.cerm.unifi.it/about-cerm/cirmmp*). The experiment has been performed at 298 K on a Bruker Avance III 800 MHz NMR spectrometer equipped with a cryoprobe. The delays $2T_N$, $2\tau_N$ $2\tau$, and $2\tau_{CN}$ were set to 24, 27, 9, and 25 ms, respectively. Gaussian cascade Q3 pulses [31] were used for band selective excitation and inversion along the $^{13}$C channel.

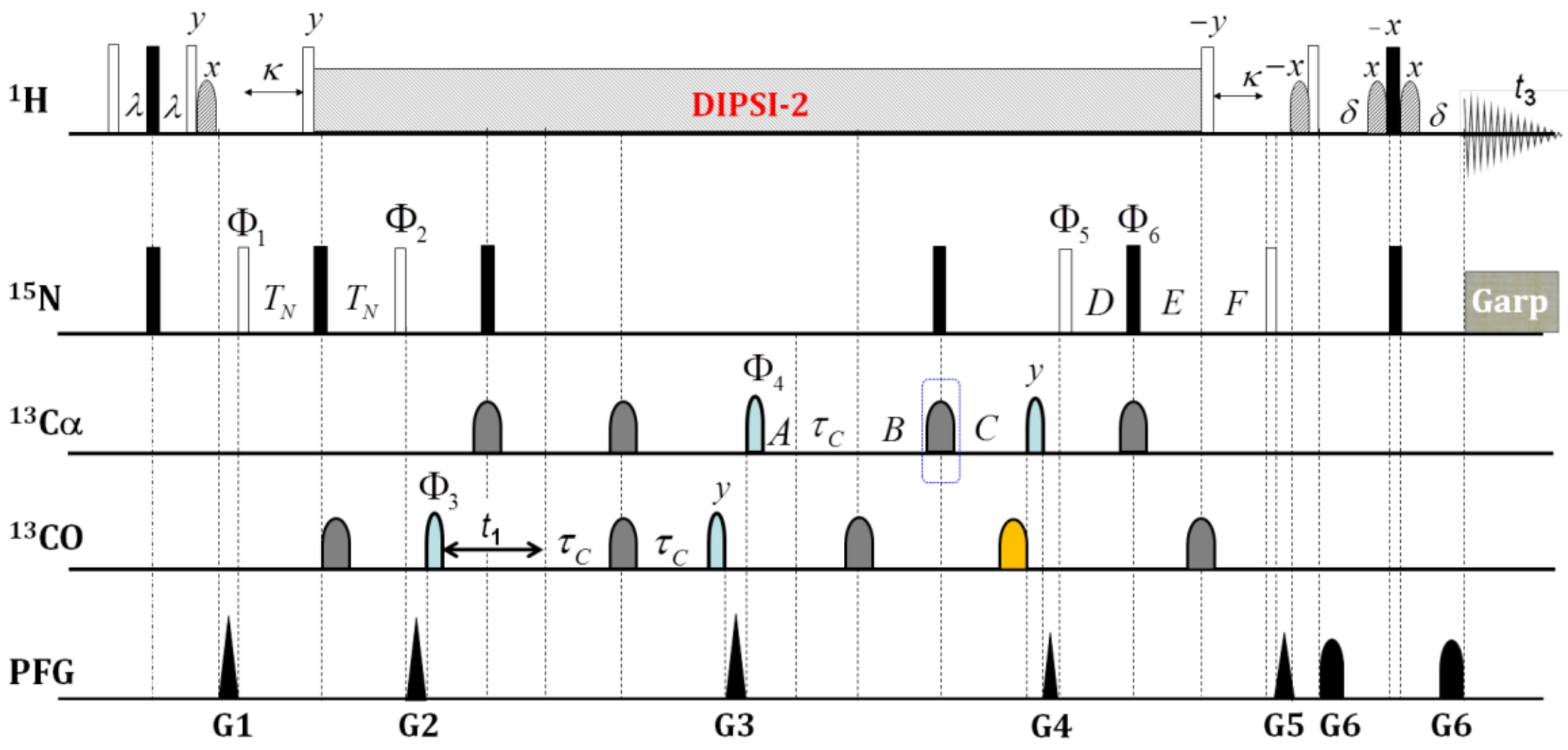

**Figure 2:** Pulse sequence for RD (4,3)D *hn*COCANH experiment. Narrow (hollow) and wide (filled black) rectangular bars represent non-selective 90° and 180° pulse, respectively. Unless indicated, the pulses are applied with phase $x$. Proton decoupling using the waltz-16 decoupling sequence [32,33] with field strength of 6.3 kHz is applied during most of the $t_1$ ($^{13}$C)and $t_2$ ($^{15}$N) evolution periods, and $^{15}$N decoupling using the GARP-1 sequence [34] with the field strength 0.9 kHz is applied during acquisition. Standard Gaussian cascade pulses [31] –shape Q3 for 180° inversion (filled black and grey; width 200 μs) and shape Q5 for 90° excitation (hollow, width 310 μs)– were used along the carbon channel. The strength of the $^{13}$C$^{\alpha}$ pulses is adjusted so that they cause minimal excitation of carbonyl carbons and that of 180° $^{13}$C' shaped pulse so that they cause minimal excitation of $^{13}$C$^{\alpha}$. The blue encircled pulse on $^{13}$C$^{\alpha}$ channel is crucial for tuning the experiment for generation of different check points in the final spectrum (like glycines/alanines and serines/threonines, for detail see the text). The values for the individual periods containing $t_1$ evolution of $^{13}$C$^{\alpha}$ nuclei are: A = $t_1/2$, B = $\tau_{CN} - \tau_C$, and C = $\tau_{CN} - t_1/2$. The $^{13}$C' nuclei are co-evolved with for $^{13}$C$^{\alpha}$ nuclei, but in the real time manner. The values for the individual periods containing $t_2$ evolution of $^{15}$N nuclei are: D = $\tau_N - t_2/2$, E = $\tau_N$, and F = $t_2/2$. The other delays are set to $\lambda$ = 2.5 ms, $\kappa$ = 5.4 ms, $\delta$ = 2.5 ms, $\tau_C$ = 4.5 ms, $T_N$ = 12 ms, $\tau_{CN}$ = 12.5 ms and $\tau_N$ = 13.5 ms. The $\tau_{CN}$ must be optimized and is around 12-15 ms. The phase cycling for the experiment is $\Phi_1$ = 2($x$), 2(-$x$); $\Phi_2 = \Phi_3 = x, -x$; $\Phi_4 = \Phi_5 = x$; $\Phi_6$ = 4($x$), 4(-$x$); and $\Phi_{receiver}$ = 2($x$), 4(-$x$), 2($x$). The frequency discrimination in $t_1$ and $t_2$ has been achieved using States-TPPI phase cycling [35] of $\Phi_4$ and $\Phi_5$, respectively, along with the receiver phase. The gradient (sine-bell shaped; 1 ms) levels are as follows: $G_1$=30%, $G_2$=30%, $G_3$=30%, $G_4$=30%, $G_5$=50%, and $G_6$=80% of the maximum strength 53 G/cm in the z-direction. The recovery time after each gradient pulse was 160 μs. Before detection, WATERGATE sequence [36] has been employed for better water suppression.

A total of 1024 complex data points were collected along the direct dimension, 32 complex points were used along the indirect $F_2$ ($^{15}$N) dimension, while 96 complex points were used along the indirect $F_1$ ($^{13}$C) dimension. For each increment, 8 scans were accumulated. The recycling delay was set to 0.8 sec. The acquisition time was approximately 16 hours 10 min. The $^{1}$H, $^{15}$N, $^{13}$C$^{\alpha}$ and $^{13}$C' carrier frequencies are set at 4.7 ppm (water), 117.0 ppm, 54.0 ppm, and 188.0 ppm, respectively. The spectral widths along $^{1}$H and $^{15}$N dimensions were 10.0 and 27.0 ppm, respectively. The $^{13}$C carrier frequency for pulses in $^{13}$C$^{\alpha}$ and $^{13}$C' channel were set at 54.0 ppm (spectral width along $^{13}$C$^{\alpha}$ dimension was 40 ppm) and 174 ppm, respectively. The NMR data was processed using Topspin (BRUKER, http://www.bruker.com/) and analyzed using CARA [37].

## Results and Discussion:

### Pulse sequence and Magnetization Transfer in (4,3)D-hnCOCANH experiment

The pulse sequence for (4,3)D-hnCOCANH has been shown in **Fig. 2**. It has been derived by simple modification of the basic 3D HN(C)N experiment described earlier [38] and differ in the way the $t_1$ evolution is handled according to reduced dimensionality (RD) NMR [23,24]. Therefore, the transfer efficiency functions which dictate the intensities of self and sequential correlation peaks in (4,3)D-hnCOCANH spectrum would be the same as those in the HN(C)N spectrum described previously [38-40] except that $^{13}C'$ and $^{13}C^{\alpha}$ chemical shifts are co-evolved during $t_1$ evolution instead of amide $^{15}N$ chemical shifts. Therefore, only the distinguishing features of these spectra have been presented here. **Figure 3A and 3B** traces the magnetization transfer pathway along with the respective frequency labeling schemes in the (4,3)D-hnCOCANH pulse sequence. As shown schematically in **Figure 3C**, the peaks appear at the following coordinates in the (4,3)D-hnCOCANH spectrum:

$$F_1 = C^+_{i-1}/C^-_{i-1},\ (F_2,F_3)\ = (H_i,N_i),(H_{i-1},N_{i-1})$$
$$F_2 = N_i,\ (F_1,F_3) = (H_i,C^+_i),(H_i,C^+_{i-1}),(H_i,C^-_i),(H_i,C^-_{i-1})$$

The letters "H" and "N" here refer to amide $^1H$ and $^{15}N$ chemical shifts, whereas the letters "$C^+$" and "$C^-$" refer to the linear combination of $^{13}C^{\alpha}$ and $^{13}C'$ chemical shifts and have been represented here as $^{13}C^+$ and $^{13}C^-$, respectively. Depending upon the offset of RF pulses used along $^{13}C'$ channel (i.e. $^{13}C'_{offset}$), these can be evaluated according to reduced dimensionality NMR concept [23] as:

$$^{13}C^+ = {}^{13}C^{\alpha}_{obs} + ({}^{13}C'_{obs} - {}^{13}C'_{offset}) \qquad (1)$$
$$^{13}C^- = {}^{13}C^{\alpha}_{obs} - ({}^{13}C'_{obs} - {}^{13}C'_{offset}) \qquad (2)$$

Thus, the $F_2(^{15}N)$-$F_3(^1H)$ plane corresponding to $F_1 = C^+_{i-1}/C^-_{i-1}$, shows a self ($H_i$, $N_i$) and a sequential ($H_{i-1}$, $N_{i-1}$) amide correlation peak and the $F_1(^{13}C)$-$F_3(^1H)$ plane corresponding to $F_2 = N_i$, shows two inter-residue ($H_i,C^+_{i-1}$ and $H_i,C^-_{i-1}$) and two intra-residue ($H_i,C^+_i$ and $H_i,C^-_i$) correlation peaks **(Fig. 3C)**. Frequency selection along all the indirect dimensions has been done by States-TPPI method [35] and the data is processed using normal Fourier Transformation (FT) like standard NMR experiments.

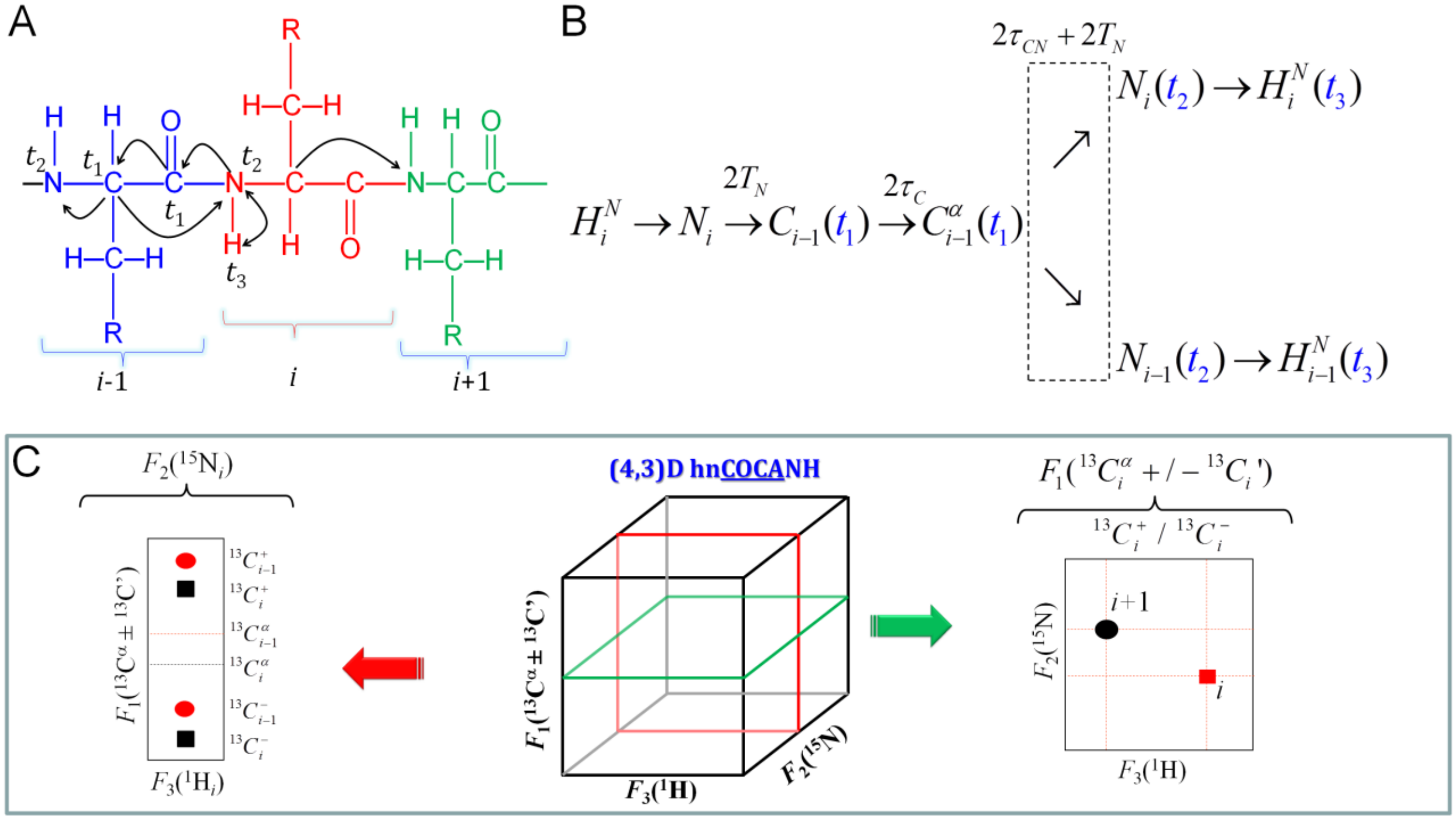

**Figure 3:** **(A)** and **(B)** Schematic illustrations, respectively, of the magnetization transfer pathway and the selected coherence transfer pathways employed in reduced dimensionality (4,3)D- hnCOCANH experiment. The magnetization flow from $H^N(i)$ is shown and the frequency labeling of the appropriate nuclei are indicated. In **(B)**, $2T_N$, $2\tau_C$, $2\tau_{CN}$ and $2\tau_N$ are the delays during which the transfers indicated by the arrows take place in the pulse sequence. **(C)** Schematic representation of the three-dimensional (4,3)D hnCOCANH spectrum. The correlations observed in the $F_1(^{13}C)$–$F_3(^1H)$ plane at the $^{15}N$ chemical shift of residue *i* are shown on left side and the correlations observed in the $F_2(^{15}N)$–$F_3(^1H)$ plane at the $^{13}C^+/^{13}C^-$ chemical shift of residue *i* (where $^{13}C^+ = {}^{13}C^{\alpha} + {}^{13}C'$ and $^{13}C^- = {}^{13}C^{\alpha} - {}^{13}C'$) are shown on the right side. Squares and circles represent the self (intra-residue) and sequential (inter-residue) peaks, respectively. Red and black represent positive and negative phase of peaks, respectively.

### Amino Acid Sequence-Dependent Peak Patterns

Like the basic HN(C)N experiment, the inter- and intra-residue correlation peaks in different planes of the (4,3)D-hnCOCANH spectrum have the opposite peak signs except in special situations. Considering different triplets of residues, covering the general and all the special situations, the expected peak patterns in different planes of the (4,3)D-hnCOCANH spectrum have been shown in **Figure 4** where inter- and intra-residue correlation peaks have been shown by circles and squares, respectively. **Figure 4A** shows the expected peak patterns (for the example stretch -PZXGYP- covering all the various possibilities) in the $F_2(^{15}N)$-$F_3(^1H)$ planes of the 3D spectrum at the $F_1(^{13}C^+/^{13}C^-)$ chemical shift of residue, *i*-1 (where $^{13}C^+ = {}^{13}C^{\alpha} + {}^{13}C'$ and $^{13}C^- = {}^{13}C^{\alpha} - {}^{13}C'$). **Figure 4B** shows the expected peak patterns (for the triplet stretches shown above each panel) in the $F_1(^{13}C)$-$F_3(^1H)$ planes of the spectrum at the $F_2(^{15}N)$ chemical shift of the central residue. The filled and empty circles/squares represent positive and negative peaks, respectively. The actual signs in the spectrum are dictated by whether the *i*−1 residue is a glycine or otherwise, and of course by the phasing of the spectrum. Here, we have made the inter-residue correlation peak negative for a triplet sequence –XYZ-, where X, Y, and Z are any residues other than glycines and prolines which have been represented here by letters "G" and "P", respectively. Accordingly the peak patterns have been shown in **Figure 4**.

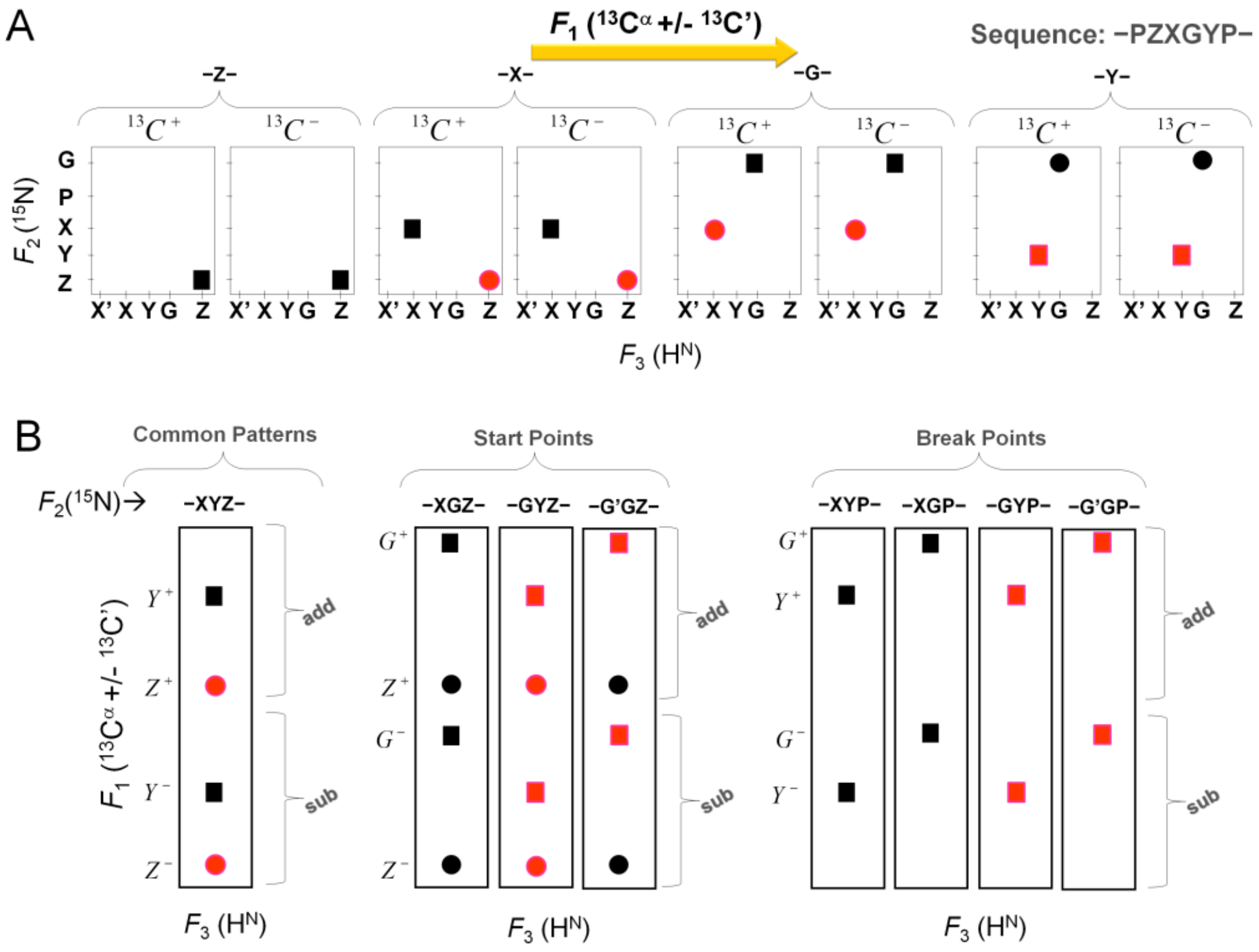

**Figure 4:** **(A)** Schematic peak patterns in the $F_2$($^{15}$N)-$F_3$($^1$H) planes of the RD 3D-hnCOCANH spectrum at the $F_1$($^{13}C^+$/$^{13}C^-$) chemical shift of residue, $i$-1 (where $^{13}C^+$ = $^{13}C^{\alpha}$ + $^{13}C'$ and $^{13}C^-$ = $^{13}C^{\alpha}$ – $^{13}C'$). An arbitrary amino acid sequence (covering all the common and special peak patterns) is chosen to illustrate the triplet specific peak patterns in these planes of the spectra. X, Y and Z are any residues other than glycine and proline. G and P, respectively, represent the glycine and proline residues. Black and red colors represent positive and negative signs, respectively. Squares and circles represent the diagonal and sequential peaks, respectively. **(B)** Schematic peak patterns in the $F_1$($^{13}$C)-$F_3$($^1$H) planes and the corresponding triplet of residues. The patterns involving G serve as start and/or check points during a sequential assignment walk through the spectrum. GGP, which is likely to be less frequent than the others, may serve as a unique start point. The patterns involving P identify break points during a sequential walk and thus they also serve as check points. The pattern, which does not involve a G or a P, is the most common one occurring through the sequential walk.

From the figure, it is clear that glycines make an important difference in the expected patterns of peaks. For glycines, the evolutions of the magnetization components are slightly different because of the absence of the $\beta$-carbon. This in turn generates some special patterns depending on whether the ($i$-1)$^{th}$ residue is a glycine or otherwise (see **Fig. 4**). These special peak patterns –which help in the identification the residues following glycines in the sequence– provide important start or check points during the course of sequential assignment process. Note that the $^{13}C^{+/-}$ chemical shifts of glycines are still most up-field shifted (like $^{13}C^{\alpha}$ shifts) compared to other amino acids **(Fig. 1)** which provides another check for identifying glycines and the residues following glycines. Prolines which do not have amide proton further give rise to special patterns in the spectrum. Prolines, at position *i*-1 in the triplet stretches, results in the absence of sequential amide correlation peak ($H_{i-1}$, $N_{i-1}$) in $F_2$($^{15}$N)-$F_3$($^{1}$H) plane of the spectrum at the $F_1$($^{13}C^{+/-}$) chemical shift of residue, *i*–1 (see **Fig. 4A**), whereas a prolines, at position *i*+1 in the triplet stretches results in the absence of inter-residue correlation peaks ($H_i, C^{+}_{i-1}$ and $H_i, C^{-}_{i-1}$) in $F_1$($^{13}$C)-$F_3$($^{1}$H) plane of the spectrum at the $F_2$($^{15}$N) chemical shift of residue, *i* (see **Fig. 4B**). These special peak patterns –which help in the identification of residues following prolines in the sequence–, provide stop or break points during the course of the sequential assignment walk. This facilitates the assignment process further.

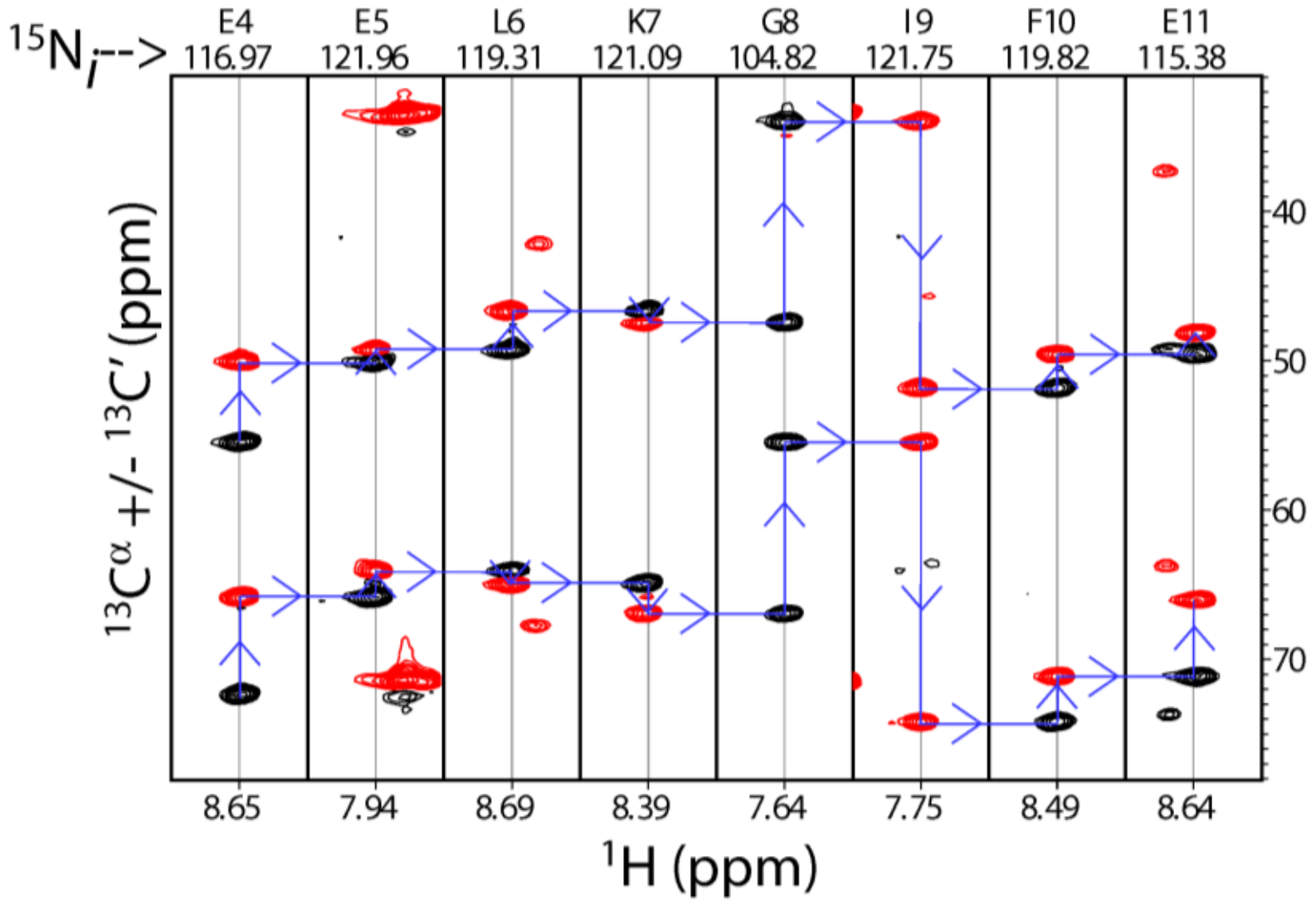

**Figure 5:** The illustrative stretch of sequential walk through the strips along $^{13}$C dimension and centered about amide $^{1}$H chemical shift of residue *i* in $F_2$($^{13}$C)-$F_3$($^{1}$H) planes of the (4,3)D-hnCOCANH spectrum of bovine apo calbindin-d9k (for residues Glu4-Glu11) at $^{15}N_i$ chemical shift. Each strip contains information about self (*i*) and sequential (*i*-1) $^{13}$C correlation peaks. The red and black contours indicate positive and negative peaks, respectively. The labels and numbers at the top in each panel identify the residue and the respective $F_1$($^{15}$N) chemical shifts. A horizontal line connects self peak (red here) in one plane to a sequential peak (black here) in the adjacent plane on the right. However, the special patterns appear for strips of glycines (self peak appears positive/black in sign) and the residues following glycines in the sequence (sequential peak appears negative/red in sign). The fact here has been illustrated using the panels of Gly8 and Ile9. These special patterns help during the sequential assignment walk through the spectrum.

The above described features of the reduced dimensionality experiment (4,3)D-hnCOCANH have been tested and demonstrated here on a small globular protein bovine apo calbindin (75 aa). **Figures 5 and 6** show the experimental demonstration of these features, respectively, for the $F_1$($^{13}$C)-$F_3$($^{1}$H) and $F_2$($^{15}$N)-$F_3$($^{1}$H) planes of the (4,3)D-hnCOCANH spectrum. A point to be mentioned here is that to achieve best separation of sum and difference frequencies (i.e. $^{13}C^{+}$ and $^{13}C^{-}$) in the final (4,3)D-hnCOCANH spectrum, the carriers of $^{13}C^{\alpha}$ and $^{13}$C' RF pulses have been placed at 54 and 188 ppm, respectively, whereas the final data is processed for the $^{13}C^{\alpha}$ offset of 54 ppm. In order to achieve this separation between the addition and subtraction frequencies (i.e. $^{13}C^{+}$ and $^{13}C^{-}$), one should be careful about (i) the spectral width used along the co-evolved dimension and (ii) the RF pulse carrier along $^{13}C'$ channel. The combined analysis of average $^{13}C^{\alpha}$ and $^{13}$C' chemical shifts of all the amino acids (**Fig. 1**; data taken from BMRB) and the **Eqns 1 and 2** revealed that for a given protein the two frequencies (i.e. $^{13}C^{+}$ and $^{13}C^{-}$) will be separated from each other (i) if the $^{13}C'$ carrier is kept 1–2 ppm away from the most down fielded shifted $^{13}C'$ chemical shift (typically close to 187–190ppm) and (ii) accordingly the spectral width along the co-evolved dimension is increased (typically between 60-80 ppm).

**Filtering of Sequentially Connected HSQC Peaks along the co-evolved carbon dimension:**

As evident from **Figure 4**, the remarkable feature of (4,3)D-hnCOCANH spectrum is that (a) it provides easy discrimination between self (intra-residue) and sequential (inter-residue) correlation peaks because of their opposite peak signs and thus no other complementary experiment is required and (b) exhibit special patterns of self and sequential peaks around glycines and prolines which serve as check points for transforming the stretches of sequential connected peaks into the final assignment. Further, the co-evolution of $^{13}$C' and $^{13}C^{\alpha}$ chemical shifts provides a dispersion better compared with the dispersion of the individual chemical shifts. The fact has been demonstrated here in **Figure 1** using average $^{13}$C' and $^{13}C^{\alpha}$ chemical shifts derived from the Biological Magnetic Resonance Bank (BMRB) database [www.bmrb.wisc.edu/ref_info] for all the 20 amino acids. As shown in the figure, the average $^{13}$C' and $^{13}C^{\alpha}$ chemical shifts individually show dispersions of ~21.5 and ~8 ppm, respectively, **(Fig. 1A and 1B)**; where 1 ppm = 200 Hz, at 800 MHz spectrometer for $^{13}$C dimension. However, the addition and subtraction of these chemical shifts ($^{13}$C' and $^{13}C^{\alpha}$) provide dispersion of ~21.5 and ~25.0 ppm, respectively **(Fig. 1C and 1D);** relatively better compared to the individual ones for efficient and rapid resonance assignment following the routine assignment procedure where two sequential residues are identified through the match of carbon chemical shifts of residues *i* and *i* - 1. The process has been demonstrated here in **Figure 5.**

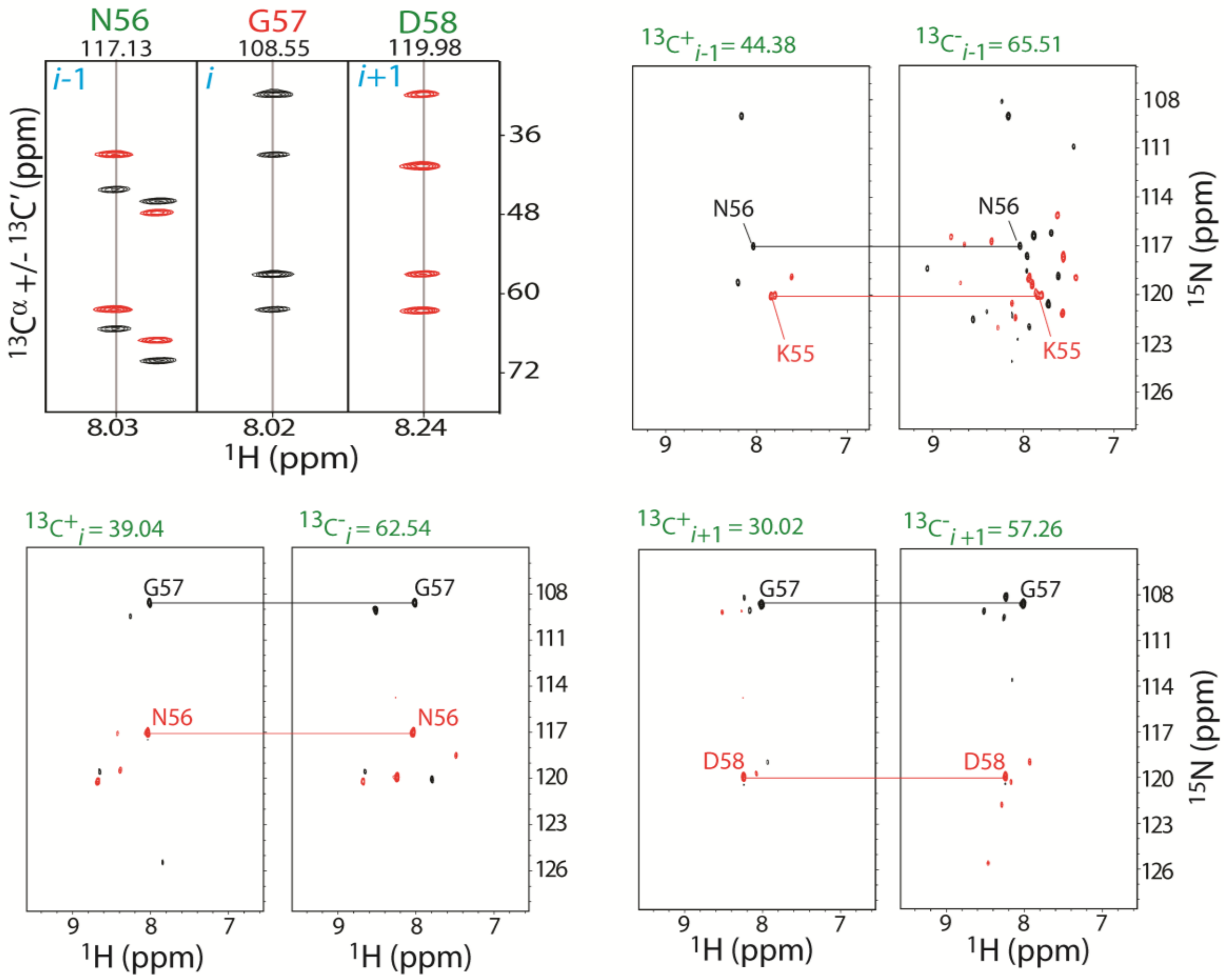

**Figure 6:** An illustrative example showing use of correlations observed in the $F_2(^{15}N)$–$F_3(^{1}H)$ planes at the $^{13}C^{+}(= {}^{13}C^{\alpha} + {}^{13}C')/{}^{13}C^{-}(= {}^{13}C^{\alpha} - {}^{13}C')$ chemical shift of residue *i* for establishing a sequential (*i* to *i*-1) connectivity between the backbone amide correlation peaks of $^{1}H$-$^{15}N$ HSQC spectrum. Red and black contours represent positive and negative phase of the peaks, respectively. The most up-field shifted $^{13}C$ correlation peaks (with reference to **Fig. 1** and **Fig. 4**) in lower (subtraction) and upper (addition) half of the spectrum help to identify the $^{13}C^{-}/{}^{13}C^{+}$ peaks for glycine which serve as start points for the sequential assignment walk. A sequential amide cross peak ($H_{i-1}$, $N_{i-1}$) should exist in both the $F_2(^{15}N)$–$F_3(^{1}H)$ planes of the spectrum at the $^{13}C^{+}/{}^{13}C^{-}$ chemical shifts of residue *i*. The other advantage is the opposite peak signs for the self ($H_i$, $N_i$) and sequential ($H_{i-1}$, $N_{i-1}$) amide correlation peaks. These spectral features help to reduce the search for the sequential amide correlations rather than making the search in various planes of the 3D spectrum (as is the case with other presently used assignment strategies).

However on top of this, the other beneficial feature is that the $^{1}H$–$^{15}N$ planes of the spectrum provide unambiguous filtering of sequentially connected HSQC peaks along the resulted well dispersed carbon dimension. In other words, a sequential *i* to *i*–1 connectivity between two HSQC peaks can be established directly on the $^{1}H$–$^{15}N$ plane of (4,3)D-hnCOCANH spectrum at $^{13}C^{+/-}$ chemical shift of residue *i*–1. The process has been illustrated experimentally in **Fig. 6**. Compared to the routine experiments (like HNCA, HN(CA)CO, CBCANH, etc), which also provide the identical filtering of sequentially connected HSQC peaks along the $^{13}$C dimension, the advantage with (4,3)D-hnCOCANH spectrum is that the sequential ($H_{i-1}$,$N_{i-1}$) HSQC peaks appear opposite in sign; thus can be differentiated directly from the self ($H_i$,$N_i$) HSQC peaks. The true sequential HSQC peak ($H_{i-1}$,$N_{i-1}$) will be present in opposite sign on both the $^1$H–$^{15}$N planes of spectrum at $^{13}C^{+/-}$ chemical shift of residue, *i*-1 and the process is eased further by the random and well dispersed nature of $^{13}C^{+/-}$ chemical shifts (along the jointly sampled carbon dimension). Overall, the spectrum provides efficient filtering of sequentially connected HSQC peaks which is particularly important while performing sequential assignment of intrinsically unstructured proteins, proteins containing repetitive amino acids (e.g.…TSAAGTTTE…) or repetitive stretches of amino acids (e.g. …QPLAGA… QPLAGA…) as well as for protein folding/unfolding studies by NMR. These proteins have always posed a great challenge for backbone resonance assignment following the conventional approaches where the poor dispersion of $^{13}C^{\alpha/\beta}$ chemical shifts (e.g. as in case of CBCANH) and their dependence on the amino-acid type result in the crowding of self and sequential HSQC peaks on $^1$H–$^{15}$N planes of the spectrum at the degenerate carbon chemical shifts.

## Assignment Protocol:

The assignment protocol based on reduced dimensionality (4,3)D hnCOCANH spectrum proposed here is basically an improvisation of that previously reported based on experiments 3D-hnCOcaNH and 3D-hncoCANH [21] and employs the simple reduced dimensionality concept [23,24] to achieve higher dispersion along indirectly detected $F_1$($^{13}$C) dimension. The improvisation is based on the fact that linear combination of $^{13}C^{\alpha}$ and $^{13}$C' chemical shifts provides a dispersion better compared to that of individual chemical shifts and thus helps to resolve the ambiguities arising because of degenerate $^{13}C^{\alpha}$ and $^{13}$C' chemical shifts. The efficacy and robustness of the protocol is a consequence of (a) unidirectional sequential amide and $^{13}$C correlations ($i \rightarrow i-1$) and (b) the distinctive peak patterns of self and sequential peaks in different planes of the 3D spectrum which enable ready identification of certain specific triplet sequences **(Fig. 4)**. Overall, the experiment commingles the assignment protocols based on HN(C)N [41] and HNCA/HN(CO)CA experiments and make the assignment process even faster, efficient and more accurate. The stepwise description of the protocol is as:

(1) Identify the backbone amide cross peaks in the $^1$H- $^{15}$N HSQC spectrum.

(2) Establish all the possible sequential connectivities between the HSQC peaks using the sequential amide and $^{13}$C correlations ($i \rightarrow i$-1) as illustrated in **Figures 5, 6, and 7**. Compared to routine assignment strategies, the present strategy does not require any complementary experiment for differentiating the self (intra-residue) and sequential (inter-residue) correlation peaks; as these appear opposite in their peak signs. Further, at times when self and sequential correlation peaks are very close to each other and if they are of the same sign, both the peaks appear as one in the routinely used 3D NMR spectra like HNCA, HN(CA)CO, and CBCANH and therefore the distinction between them becomes difficult. However, in case of (4,3)D-hnCOCANH spectrum, the positive and negative signs make this condition less possible and results in a dispersive peak shape which in turn indicates that the self and sequential peaks are lying close to each other. The cancellation of intensities can occur whenever two neighboring residues ($i$ and $i$–1) have almost identical $^{13}C^{+/-}$ chemical shifts. The latter condition can arise only if these residues have degenerate $^{13}C^{\alpha}$ and $^{13}C'$ chemical shifts which is very less possible. There is another important difference compared to the established assignment strategies based on experiments like HNCA/HN(CO)CA or HN(CA)CO/HNCO or CBCANH/CBCACONH where one has to scan the nitrogen planes to find the matching carbon peaks. This is very tedious and time taking process and may lead to ambiguities as well. However, in the present case this would not be the case because the carbon planes would provide the positions of $N_i$ and $N_{i-1}$ (**Fig 6**). More explicitly, start with an HSQC peak *i*. The $^{15}N$ plane for this would provide the $C_i^+, C_i^-, C_{i-1}^+$ and $C_{i-1}^-$ positions (**Fig 3C**). The $C_{i-1}^+$ and $C_{i-1}^-$ planes would then provide the $N_i$ (already known) and $N_{i-1}$ positions; these also appear in opposite signs and thus can easily be differentiated. Thus there is no need to scan the various $^{15}N$ planes. The process has been illustrated in **Fig. 6** for the amino acid residues Asn56-Asp58 of the bovine apo calbindin.

(3) Transform the protein primary sequence into a format like – XGBXGGBXP– where all residues are replaced by letter X except for glycines, residues following glycines, and prolines which are replaced by letters G, B and P, respectively.

(4) Use the check point information derived from triplet specific peak patterns (**Fig. 4**) to transform the stretches of sequentially connected HSQC cross peaks into a format like –XGBXGGBXP–. Here X is any HSQC peak giving common peak pattern and G is the HSQC peak identified as glycine (referred here as the start point, **Fig. 6**) and B is the HSQC peak identified as residue following glycine from triplet specific peak patterns (**Fig. 3**). P here refers to break point identified again from triplet specific peak patterns (**Fig. 4**). Now the transformed stretches of sequential connectivities (i.e. – XGBXGGBXP–) are compared with the transformed primary sequence to find a match. The process has been illustrated schematically in **Fig. 7**.

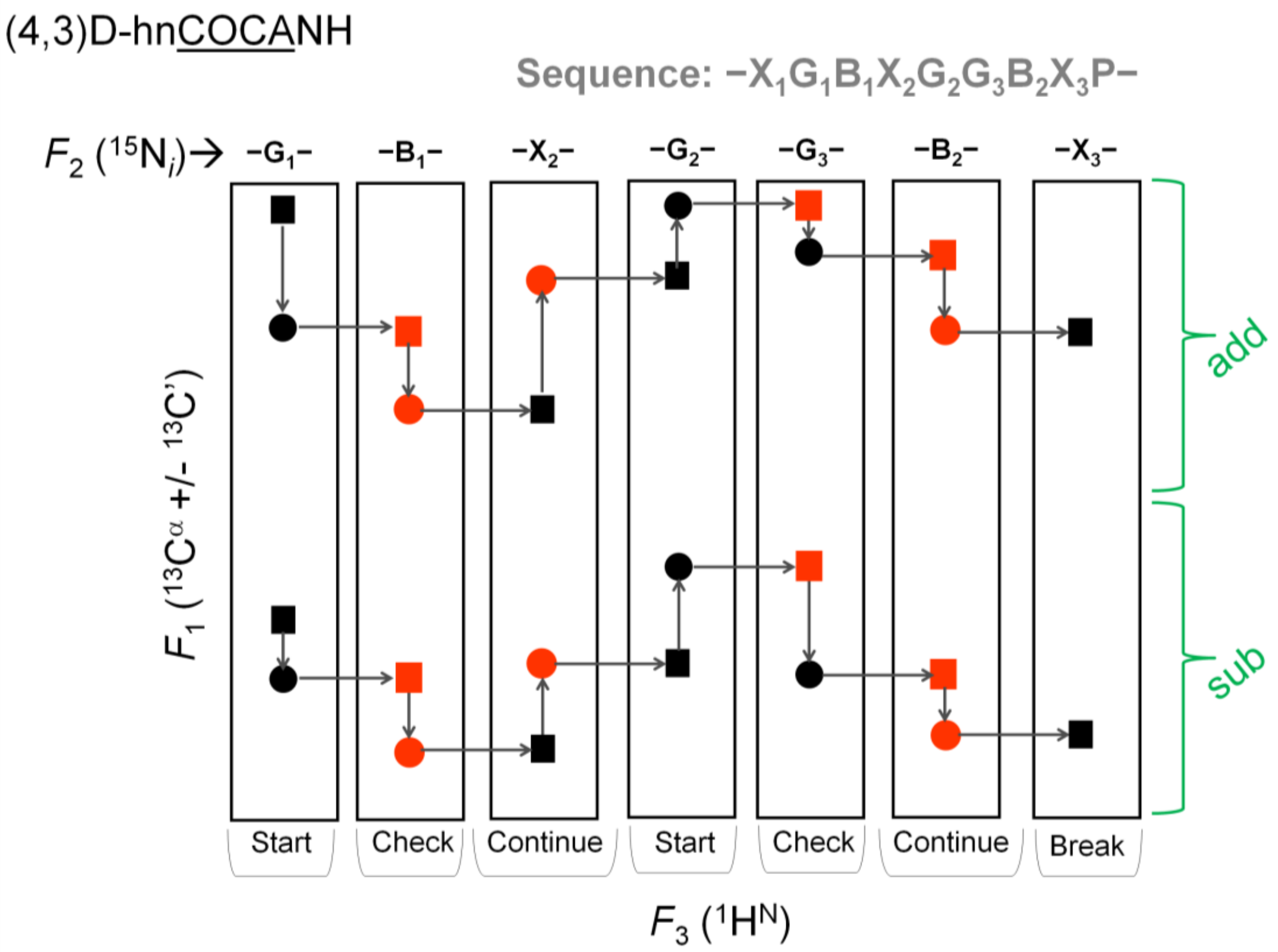

**Figure 7:** The schematic sequential assignment walk protocol through the $F_1$($^{13}$C)–$F_3$($^{1}$H) planes of the reduced dimensionality (4,3)D- hnCOCANH experiment. An arbitrary amino acid sequence is chosen to illustrate the start, continue, check, and break points during the sequential assignment walk. Squares and circles represent the self (intra-residue) and sequential (inter-residue) peaks, respectively. Red and black represent positive and negative phase of peaks, respectively. The patterns of positive and negative peaks are drawn according to the triplet specific peak patterns contained in this spectrum (see all the peak patterns in **Fig. 4**). The residue identified on the top of each strip identifies the central residue of the triplet.

Thus, the single experiment provides direction specific sequential walk and no complementary experiment is required. Even the explicit side chain assignment would not be very necessary to decide on the correctness of the sequential assignment because of the large number of various check points that are generally available. Following the above protocol, the complete sequence specific assignment of backbone ($^{1}$H, $^{15}$N, $^{13}$C$^{+}$, and $^{13}$C$^{-}$) resonances is established initially and then the backbone $^{13}$C$^{\alpha}$ and $^{13}$C' chemical shifts are calculated as:

$$^{13}C^{\alpha}_{obs} = (^{13}C^{+} + {}^{13}C^{-})/2 \qquad (3)$$

$$^{13}C'_{obs} = (^{13}C^{+} - {}^{13}C^{-})/2 \qquad (4)$$

The performance of the above protocol has successfully been tested on $^{13}$C/$^{15}$N labeled bovine apo-calbindin and the complete backbone resonance assignment was established in about 7-8 hours of manual analysis of the spectrum (**Figure 8A and Table 1**).

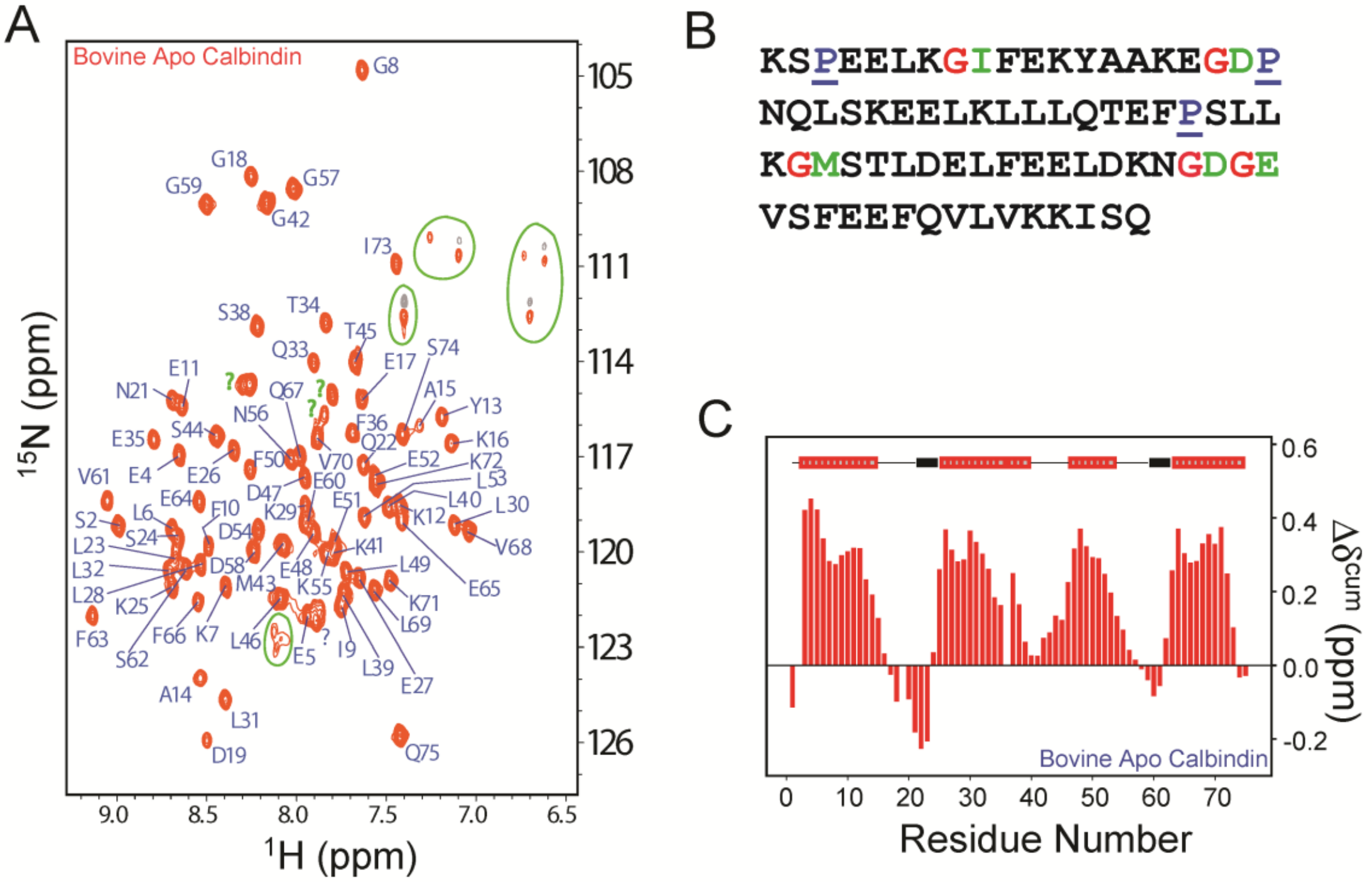

**Figure 8:** **(A)** Finger print $^{1}$H-$^{15}$N HSQC spectrum of bovine apo-calbindin-d9k showing the sequence specific assignments at pH 6.0 and 298 K. The unassigned peaks are either circled or have been marked by question mark. **(B)** The amino acid sequence of the protein (mutant P47M, residues 5-79, swissprot accession is **P02633**) with the glycines and prolines highlighted to show their distribution along the sequence. The ***glycines*** –referred here as primary check points– are highlighted in red color, ***residues following glycines*** in the sequence –referred here as secondary check points– in green color and ***prolines*** –which serve as stop points during the sequential assignment walk– in blue color and by underline. **(A)** Plot of cumulative secondary chemical shifts ( $\Delta\delta^{cum}$, from **Eqn. 5**, described in the text) determined using random coil values derived from the CamCoil Webserver (http://www-vendruscolo.ch.cam.ac.uk/camcoil.php) [42]. Native alpha-helical and beta-sheet elements (reported previously [43]: PDB ID = **1CLB**) have been marked above the plot by red and black bars, respectively.

### Tunability and Versatility of the Experiment:

A point to be mentioned here is that like the basic experiment HN(C)N, the (4,3)D-hnCOCANH experiment can also be tuned for additional amino-acid specific (i.e. alanines and serines/ threonines specific) patterns of intra-residue (self) and inter-residue (sequential) correlation peaks and hence enable identification of additional triplets of residues. This is accomplished by changing the bandwidth of the 180° inversion pulse on $^{13}C^{\alpha}$ channel during the $\tau_{CN}$ evolution period (encircled pulse in **Fig. 2**) as described previously [39,40]. Accordingly the experiment can have three variants: (a) glycine variant (described here in the present paper and designated as (4,3)D-hnCOCANH-G), (b) alanine variant (which provides identification of triplet stretches containing glycines/alanines and prolines and designated as (4,3)D-hnCOCANH-A), and (c) serine/threonine variant (which provides identification of triplet stretches containing serines/threonines and prolines and designated as (4,3)D-hnCOCANH-ST). However, to avoid the confusion, the results from normal experiment (i.e. (4,3)D-hnCOCANH-G) have been presented here. However, a point is needed to be mentioned here is that, if the protein sequence is rich in glycines and if they are very well dispersed all along the sequence (as is the case with bovine calbindin, **Fig. 8B**), the normal (4,3)D-hnCOCANH-G experiment can lead to complete assignment. However, if the protein sequence has only a few glycines and prolines and they are far removed along the sequence, the other variants of the experiment can be used to complete the backbone resonance assignment; where the selection of the variant –i.e. (4,3)D-hnCOCANH-A or (4,3)D-hnCOCANH-ST– is made depending upon the distribution of glycines/alanines and serines/threonines, respectively. Alternatively, one can also use the HSQC type experiments designed in our lab previously [44-46] for getting the additional check points (like alanines and serines/threonines).

**Estimation of secondary structure of the Protein:**

The estimation of secondary structure forms the basis for three-dimensional structure determination of proteins. The secondary structural information also provide the foundation for (i) protein folding/unfolding studies by NMR [4,5] and (ii) structural proteomics projects where the feasibility of carrying out structural studies of proteins is assessed based on its 'foldability' [47,48]. For secondary structure estimation, the method relies only on backbone $^{13}C^{\alpha}$ and $^{13}C'$ chemical shifts and its reliability has been evaluated here by calculating the cumulative secondary chemical shifts along the protein sequence as:

$$\Delta\delta^{CUM} = \frac{\Delta\delta(C^{\alpha})}{25} + \frac{\Delta\delta(C')}{10} \qquad (5)$$

where $\Delta\delta$ represents the individual secondary shift calculated as $\Delta\delta = \delta_{obs} - \delta_{rc}$; where $\delta_{obs}$ represents the observed chemical shift and $\delta_{rc}$ represents the random coil chemical shift. The sequence corrected random coil chemical shifts are derived from the *CamCoil* Web-server (http://www-vendruscolo.ch.cam.ac.uk/camcoil.php) [49]. The normalizations used for the individual secondary shifts in **Eqn. 5** are based on the span of the chemical shift of the respective nuclei in folded proteins. **Figure 8C** shows the plot of residue wise cumulative secondary chemical shifts ($\Delta\delta^{CUM}$, **Eqn. 5**) estimated for the bovine apo-calbindin (75 aa). The native secondary structural elements deduced from the NMR structural ensemble (**PDB ID = 1CLB**) are also shown at the top of the plot. The comparison clearly reveals that the secondary structural information estimated using $^{13}C^{\alpha}$ and $^{13}C'$ chemical shifts is nicely matching with that the native secondary structure of the protein (derived from **PDB ID = 1CLB**) except for some residues present in turns and loop regions connecting the secondary structural elements. Thus, the backbone ($^{1}H^{N}$, $^{15}N$, $^{13}C^{a}$ and $^{13}C'$) chemical shifts can be used for secondary structure estimation of proteins with reasonable accuracy. In other words, the experiment and based on assignment method presented here will serve as a powerful tool for secondary structure determination of proteins .

**Table 1:** Extraction of backbone $^{13}C^{\alpha}$ and $^{13}C'$ chemical shifts (in ppm) from the spectral chemical shifts ($^{13}C^{+}$ and $^{13}C^{-}$) and their transformation into the secondary structure information using the chemical shift index (CSI) method wherein, the deviation of the observed chemical shift of a given residue (here $^{13}C^{a}$ and $^{13}C'$ chemical shifts) from its random-coil value provides information on its secondary structure [25].

| Residue No. | $^{13}C^{+}$ | $^{13}C^{-}$ | $\delta_{offset}$ ($^{13}C'$) | $\delta_{obs}$ ($^{13}C^{\alpha}$) | $\delta_{obs}$ ($^{13}C'$) | | $\delta_{RC}$ ($^{13}C^{\alpha}$) | $\delta_{RC}$ ($^{13}C'$) | $\Delta\delta$ ($^{13}C)^{cum}$ | TPA-$\Delta\delta$ ($^{13}C)^{cum}$ |
|---|---|---|---|---|---|---|---|---|---|---|
| K1 | 43.63 | 68.59 | 188 | 56.11 | 175.52 | | 56.16 | 176.63 | -0.11 | -0.11 |
| **S2** | – | – | – | – | – | | – | – | – | – |
| P3 | 57.47 | 74.81 | 188 | 66.14 | 179.33 | | 62.80 | 177.04 | 0.36 | 0.42 |
| E4 | 52.11 | 68.31 | 188 | 60.21 | 179.9 | | 56.59 | 176.58 | 0.48 | 0.45 |
| E5 | 51.36 | 66.6 | 188 | 58.98 | 180.38 | | 56.26 | 176.33 | 0.51 | 0.42 |
| L6 | 48.75 | 67.39 | 188 | 58.07 | 178.68 | | 55.05 | 177.17 | 0.27 | 0.34 |
| K7 | 49.5 | 69.42 | 188 | 59.46 | 178.04 | | 56.27 | 176.88 | 0.24 | 0.28 |
| G8 | 36.1 | 57.98 | 188 | 47.04 | 177.06 | | 45.87 | 174.23 | 0.33 | 0.27 |
| I9 | 53.91 | 76.65 | 188 | 65.28 | 176.63 | | 61.14 | 175.83 | 0.25 | 0.30 |
| F10 | 51.61 | 73.57 | 188 | 62.59 | 177.02 | | 57.63 | 175.85 | 0.32 | 0.31 |
| E11 | 50.16 | 68.44 | 188 | 59.3 | 178.86 | | 56.41 | 176.38 | 0.36 | 0.32 |
| K12 | 48.99 | 68.91 | 188 | 58.95 | 178.04 | | 55.99 | 176.51 | 0.27 | 0.32 |
| Y13 | 50.14 | 70.92 | 188 | 60.53 | 177.61 | | 57.64 | 175.64 | 0.31 | 0.23 |
| A14 | 44.2 | 64.12 | 188 | 54.16 | 178.04 | | 52.73 | 177.51 | 0.11 | 0.19 |
| A15 | 43.85 | 62.13 | 188 | 52.99 | 178.86 | | 52.56 | 177.57 | 0.15 | 0.13 |
| K16 | 47.75 | 70.27 | 188 | 59.01 | 176.74 | | 56.18 | 176.64 | 0.12 | 0.03 |
| E17 | 42.67 | 67.91 | 188 | 55.29 | 175.38 | | 56.48 | 176.69 | -0.18 | -0.02 |
| G18 | 31.73 | 59.53 | 188 | 45.63 | 174.1 | | 45.87 | 174.17 | -0.02 | -0.10 |
| **D19** | – | – | – | – | – | | – | – | – | – |
| P20 | 51.214 | 76.166 | 188 | 63.69 | 175.524 | | 62.94 | 176.80 | -0.10 | -0.09 |
| N21 | 39.32 | 66.2 | 188 | 52.76 | 174.56 | | 52.98 | 175.31 | -0.08 | -0.18 |
| Q22 | 39.48 | 69.32 | 188 | 54.4 | 173.08 | | 55.92 | 176.10 | -0.36 | -0.23 |
| L23 | 40.87 | 64.65 | 188 | 52.76 | 176.11 | | 55.12 | 177.45 | -0.23 | -0.21 |
| S24 | 44.26 | 70.18 | 188 | 57.22 | 175.04 | | 58.20 | 174.88 | -0.02 | 0.03 |
| K25 | 51.1 | 70.34 | 188 | 60.72 | 178.38 | | 56.23 | 176.64 | 0.35 | 0.26 |
| E26 | 51.45 | 67.91 | 188 | 59.68 | 179.77 | | 56.51 | 176.58 | 0.45 | 0.37 |
| E27 | 49.03 | 68.23 | 188 | 58.63 | 178.4 | | 56.27 | 176.33 | 0.30 | 0.31 |
| L28 | 47.58 | 67.46 | 188 | 57.52 | 178.06 | | 55.05 | 177.17 | 0.19 | 0.28 |
| K29 | 51.07 | 70.95 | 188 | 61.01 | 178.06 | | 56.05 | 176.54 | 0.35 | 0.29 |
| L30 | 49.47 | 66.17 | 188 | 57.82 | 179.65 | | 55.34 | 177.45 | 0.32 | 0.36 |
| L31 | 51.49 | 67.41 | 188 | 59.45 | 180.04 | | 55.41 | 177.45 | 0.42 | 0.33 |
| L32 | 49.34 | 68.52 | 188 | 58.93 | 178.41 | | 55.10 | 177.35 | 0.26 | 0.30 |
| Q33 | 47.75 | 69.23 | 188 | 58.49 | 177.26 | | 55.75 | 176.12 | 0.22 | 0.26 |
| T34 | 53.78 | 77.62 | 188 | 65.7 | 176.08 | | 61.28 | 174.87 | 0.30 | 0.18 |
| E35 | 44.77 | 67.59 | 188 | 56.18 | 176.59 | | 56.34 | 176.28 | 0.02 | 0.16 |
| **F36** | – | – | – | – | – | | – | – | – | – |
| P37 | 56.89 | 74.21 | 188 | 65.55 | 179.34 | | 62.71 | 177.14 | 0.33 | 0.25 |
| S38 | 47.73 | 72.51 | 188 | 60.12 | 175.61 | | 58.11 | 174.79 | 0.16 | 0.16 |
| L39 | 44.92 | 66.26 | 188 | 55.59 | 177.33 | | 55.40 | 177.45 | 0.00 | 0.06 |
| L40 | 44.48 | 65.5 | 188 | 54.99 | 177.49 | | 55.12 | 177.17 | 0.03 | 0.03 |
| K41 | 46.15 | 67.85 | 188 | 57 | 177.15 | | 56.28 | 176.88 | 0.06 | 0.03 |
| G42 | 31.71 | 58.95 | 188 | 45.33 | 174.38 | | 45.78 | 174.28 | -0.01 | 0.07 |
| M43 | 45.2 | 67.02 | 188 | 56.11 | 177.09 | | 55.44 | 175.67 | 0.17 | 0.09 |
| S44 | 47.04 | 72.6 | 188 | 59.82 | 175.22 | | 58.22 | 174.85 | 0.10 | 0.13 |
| T45 | 50.2 | 75.4 | 188 | 62.8 | 175.4 | | 61.48 | 174.63 | 0.13 | 0.12 |
| L46 | 47.04 | 67.56 | 188 | 57.3 | 177.74 | | 55.20 | 177.26 | 0.13 | 0.22 |
| D47 | 48.88 | 66.76 | 188 | 57.82 | 179.06 | | 54.22 | 176.39 | 0.41 | 0.32 |
| E48 | 50.43 | 68.03 | 188 | 59.23 | 179.2 | | 56.33 | 176.33 | 0.40 | 0.37 |
| L49 | 48.48 | 66.26 | 188 | 57.37 | 179.11 | | 55.13 | 177.07 | 0.29 | 0.32 |
| F50 | 49.49 | 70.15 | 188 | 59.82 | 177.67 | | 57.69 | 175.85 | 0.27 | 0.29 |
| E51 | 49.97 | 68.79 | 188 | 59.38 | 178.59 | | 56.51 | 176.58 | 0.32 | 0.29 |
| E52 | 48.68 | 68 | 188 | 58.34 | 178.34 | | 56.27 | 176.33 | 0.28 | 0.25 |
| L53 | 46.44 | 66.08 | 188 | 56.26 | 178.18 | | 55.12 | 177.26 | 0.14 | 0.20 |
| D54 | 45.08 | 66.1 | 188 | 55.59 | 177.49 | | 54.22 | 176.39 | 0.16 | 0.13 |
| K55 | 46.37 | 67.93 | 188 | 57.15 | 177.22 | | 56.49 | 176.55 | 0.09 | 0.10 |
| N56 | 41.19 | 65.07 | 188 | 53.13 | 176.06 | | 53.20 | 175.76 | 0.03 | 0.04 |
| G57 | 32.04 | 59.66 | 188 | 45.85 | 174.19 | | 45.91 | 174.17 | 0.00 | 0.01 |
| D58 | 42.93 | 65.27 | 188 | 54.1 | 176.83 | | 54.26 | 176.60 | 0.02 | -0.01 |
| G59 | 31.22 | 60.32 | 188 | 45.77 | 173.45 | | 45.74 | 173.91 | -0.04 | -0.04 |
| E60 | 43.07 | 67.81 | 188 | 55.44 | 175.63 | | 56.08 | 176.27 | -0.09 | -0.08 |
| V61 | 48.3 | 73.72 | 188 | 61.01 | 175.29 | | 61.82 | 176.10 | -0.11 | -0.05 |
| S62 | 44.11 | 69.59 | 188 | 56.85 | 175.26 | | 57.94 | 174.43 | 0.04 | 0.07 |
| F63 | 50.5 | 71.82 | 188 | 61.16 | 177.34 | | 57.71 | 175.85 | 0.29 | 0.26 |
| E64 | 51.79 | 68.89 | 188 | 60.34 | 179.45 | | 56.51 | 176.58 | 0.44 | 0.37 |
| E65 | 49.92 | 67.8 | 188 | 58.86 | 179.06 | | 56.32 | 176.28 | 0.38 | 0.30 |
| F66 | 48.19 | 72.79 | 188 | 60.49 | 175.7 | | 57.67 | 175.96 | 0.09 | 0.28 |
| Q67 | 50.01 | 69.79 | 188 | 59.9 | 178.11 | | 55.66 | 176.17 | 0.36 | 0.28 |
| V68 | 56.59 | 76.29 | 188 | 66.44 | 178.15 | | 62.08 | 175.96 | 0.39 | 0.33 |
| L69 | 48.77 | 67.91 | 188 | 58.34 | 178.43 | | 55.24 | 177.38 | 0.23 | 0.35 |
| V70 | 56.7 | 75.58 | 188 | 66.14 | 178.56 | | 61.88 | 175.86 | 0.44 | 0.33 |
| K71 | 49.99 | 68.91 | 188 | 59.45 | 178.54 | | 56.14 | 176.71 | 0.32 | 0.37 |
| K72 | 49.91 | 68.11 | 188 | 59.01 | 178.9 | | 55.87 | 176.50 | 0.37 | 0.25 |
| I73 | 50.34 | 74.36 | 188 | 62.35 | 175.99 | | 60.53 | 176.12 | 0.06 | 0.10 |
| S74 | 44.12 | 73.74 | 188 | 58.93 | 173.19 | | 58.00 | 174.80 | -0.12 | -0.03 |
| **Q75** | – | – | – | – | – | | – | – | – | – |

Note:

1. RC stands for random coil shifts derived from CamCoil Server (**http://www-vendruscolo.ch.cam.ac.uk/camcoil.php**) **[42]**.
2. $\delta_{obs}(^{13}C^{\alpha}) = [\delta_{obs}(^{13}C^{+}) + \delta_{obs}(^{13}C^{-})]/2$
3. $\delta_{obs}(^{13}C') = \delta_{offset}(^{13}C') - [\delta_{obs}(^{13}C^{-}) - \delta_{obs}(^{13}C^{\alpha})]$
4. $\Delta\delta(^{13}C)^{cum} = [\delta_{obs}(^{13}C^{\alpha}) - \delta_{RC}(^{13}C^{\alpha})]/25 + [\delta_{obs}(^{13}C') - \delta_{RC}(^{13}C')]/10$ ; is called as cumulative secondary shifts
5. TPA stands for three point averaging

## Concluding Remarks:

In conclusion, an efficient standalone method for high-throughput assignment of backbone ($^{1}H^{N}$, $^{15}N$, $^{13}C^{\alpha}$, and $^{13}C'$) resonances and secondary structure determination of proteins has been presented. Compared to currently available strategies, the method employs a single reduced dimensionality (RD) 3D NMR experiment -i.e. (4,3)D-hnCOCANH- and provides the complete sequence specific assignment of backbone ($^{1}H$, $^{15}N$, $^{13}C^{\alpha}$ and $^{13}C'$) resonances. The performance of the experiment and the utility of the method has been demonstrated here on a $^{13}C/^{15}N$ labeled bovine apo calbindin-D9K and the whole exercise (including data collection and data analysis) has been completed in a day.

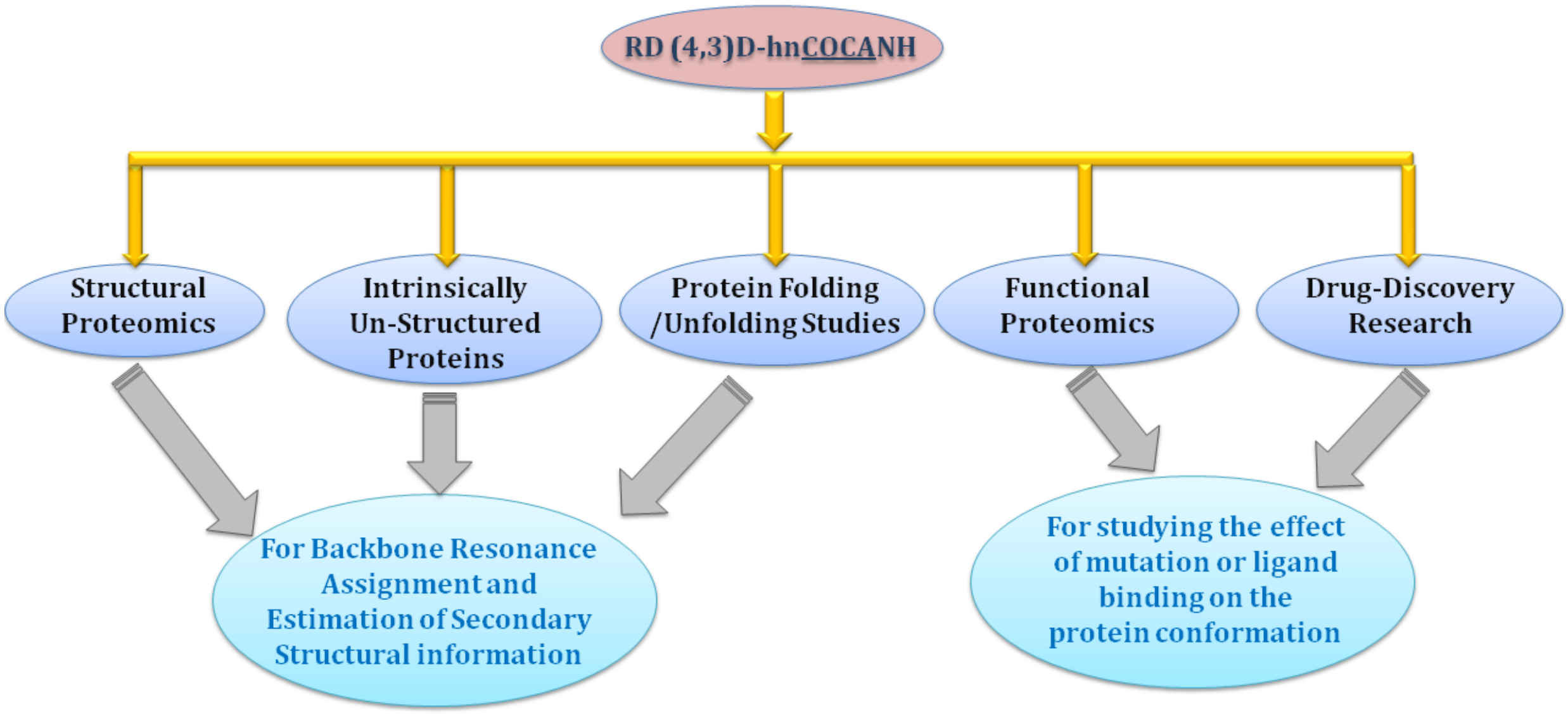

**Figure 9:** Schematic showing the range of applications of the standalone assignment method presented here based on the single reduced dimensionality (4,3)D- hnCOCANH experiment.

Overall, the method has its immense value in NMR investigations of both folded and functional unfolded proteins because of its efficacy, speed and simplicity. The wide range of applications of the method have been summarized briefly in **Fig. 9**. The method is very well suited for automated data analysis as well where keeping the total number of spectra small is highly desirable. The use of single experiment further helps to get rid of the complications arising because of inter-spectral variations of chemical shifts; the problem is common in proteins which are unstable in solution or tend to precipitate in matter of days.

The method will also serve as a valuable NMR assignment method in drug discovery research programs especially for SAR by NMR (i.e. studying Structure Activity Relations by NMR) [9,50]. It is the most commonly used method to verify binding, stoichiometry, and identification of the binding site on the protein targets. In practice, SAR by NMR exploits the differences in amide chemical shifts in the 2D $^{1}H$-$^{15}N$

HSQC spectra of target protein in its free and ligand bound form; provided that the backbone amide chemical shifts are all assigned for the protein targets both in its free and ligand/drug bound form. This is done thousands of times to find the best drug candidate. However sometimes, the chemical shift changes induced by the complex formation are so large that the sequential assignment for amide resonances may lost. Further, when the high-affinity ligands are in a slow exchange with their target receptors, there will be two sets of peaks and the identity of shifted resonances of the complex cannot be unraveled simply by titration [51]. In such a case, the presented reduced dimensionality experiment will be the best choice simultaneously to re-establish the lost sequential assignments of target proteins (both in its free and ligand/drug bound states) as well as to probe the conformational changes induced upon ligand binding. Similarly, the method can also be used to re-establish the lost resonance assignment of protein mutants (specific mutations for folding and/or function) and to probe the conformational changes induced by mutation to ascertain the role of specific residues in a particular function.

Finally, a few comments are required to be mentioned about the sensitivity of the experiment. This is important as the experiment employs many steps of magnetization transfer via several long transfer delays **(Fig. 2)** rendering low sensitivity to the experiment. However with the currently available higher magnetic fields and efficient cryo-probes offering higher signal to noise ratio, the sensitivity of the experiment will be not be a serious issue and it can be successfully applied on higher (< 15 kDa) molecular weight proteins. Since the experiment does not involve aliphatic protons, the sensitivity problem can be circumvented further by incorporation of the TROSY option [52] along with per-deuteration of the proteins to reduce relaxation losses. Moreover, to save on experiment time, the experiment can be modified according to recently introduced (i) BEST NMR [53], (ii) L-optimization [54], or (iii) non-uniform sampling [55] approaches for rapid data collection. Taken together, the standalone method presented here will be an important backbone assignment tool for structural and functional proteomics, protein folding, and drug discovery research programs by NMR.

## Reference:

[1] A. Yee, A. Gutmanas, C.H. Arrowsmith, Solution NMR in structural genomics, Curr. Opin. Struct. Biol. 16 (2006) 611-617.

[2] K. Wuthrich, NMR of Proteins and Nucleic Acids, (1986)

[3] G. Wagner, An account of NMR in structural biology, Nat. Struct. Biol. 4 Suppl (1997) 841-844.

[4] H.J. Dyson, P.E. Wright, Unfolded proteins and protein folding studied by NMR, Chem. Rev. 104 (2004) 3607-3622.

[5] H.J. Dyson, P.E. Wright, Elucidation of the protein folding landscape by NMR, Methods Enzymol. 394 (2005) 299-321.

[6] DeLucas L J, Brouillette C G, Ramnarayan K, Mylvaganam S, Tools for Structural Genomics accelerating the structure pipeline, Drug Discovery World Summer 2004: Structural Genomics (2004) 65-74.

[7] H.J. Dyson, P.E. Wright, Intrinsically unstructured proteins and their functions, Nat. Rev. Mol. Cell Biol. 6 (2005) 197-208.

[8] M. Pellecchia, D.S. Sem, K. Wuthrich, Nmr in drug discovery, Nat Rev Drug Discov 1 (2002) 211-219.

[9] R. Powers, Applications of NMR to structure-based drug design in structural genomics, J Struct Func Genom 2 (2002) 113-123.

[10] D.A. Erlanson, J.A. Wells, A.C. Braisted, Tethering: fragment-based drug discovery, Annu. Rev Biophys. Biomol. Struct 33 (2004) 199-223.

[11] D.A. Erlanson, R.S. McDowell, T. O'Brien, Fragment-based drug discovery, J Med. Chem 47 (2004) 3463-3482.

[12] Cavanagh J, Fairbrother W J, Palmer A G, Skelton N J, Protein NMR Spectroscopy: Principles and Practice, (2006)

[13] P. Permi, A. Annila, Coherence transfer in proteins, Progress in Nuclear Magnetic Resonance Spectroscopy 44 (2004) 97-137.

[14] Sattler M, Schleucher J, Griesinger C, Heteronuclear multidimensional NMR experiments for the structure determination of proteins in solution employing pulsed field gradients, Progress in Nuclear Magnetic Resonance Spectroscopy 34 (1999) 93-158.

[15] S. Grzesiek, A. Bax, Improved 3D triple-resonance NMR techniques applied to a 31 kDa protein, J. Magn Reson. A 96 (1992) 432-440.

[16] L.E. Kay, M. Ikura, R. Tschudin, A. Bax, Three-Dimensional Triple-Resonance NMR Spectroscopy of Isotopically Enriched Proteins, J. Magn Reson. B 89 (1990) 496-514.

[17] R.T. Clubb, V. Thanabal, G. Wagner, A constant-time three-dimensional triple-resonance pulse scheme to correlate intraresidue $^{1}H^{N}$, $^{15}N$, and $^{13}CO$ chemical shifts in $^{15}N$---$^{13}C$-labelled proteins, J. Magn Reson. A 97 (1992) 213-217.

[18] S. Grzesiek, A. Bax, An Efficient Experiment for Sequential Backbone Assignment of Medium-Sized Isotopically Enriched Proteins, J. Magn Reson. B 99 (1992) 201-207.

[19] S. Grzesiek, A. Bax, Correlating Backbone Amide and Side Chain Resonances in Larger Proteins by Multiple Relayed Triple Resonance NMR, J. Am. Chem Soc 114 (1992) 6293.

[20] N.S. Bhavesh, S.C. Panchal, R.V. Hosur, An efficient high-throughput resonance assignment procedure for structural genomics and protein folding research by NMR, Biochemistry 40 (2001) 14727-14735.

[21] D. Kumar, J.G. Reddy, R.V. Hosur, hnCOcaNH and hncoCANH pulse sequences for rapid and unambiguous backbone assignment in (13C, 15N) labeled proteins, J. Magn Reson. 206 (2010) 134-138.

[22] BMRB, Statistics Calculated for Selected Chemical Shifts from Atoms in the 20 Common Amino Acids (http://www.bmrb.wisc.edu/ref_info/), (2011)

[23] T. Szyperski, G. Wider, J.H. Bushweller, K. Wuthrich, Reduced dimensionality in triple-resonance NMR experiments, J. Am. Chem Soc 115 (1993) 9307-9308.

[24] T. Szyperski, D.C. Yeh, D.K. Sukumaran, H.N. Moseley, G.T. Montelione, Reduced-dimensionality NMR spectroscopy for high-throughput protein resonance assignment, Proc. Natl. Acad. Sci. U. S. A 99 (2002) 8009-8014.

[25] D.S. Wishart, B.D. Sykes, The 13C chemical-shift index: a simple method for the identification of protein secondary structure using 13C chemical-shift data, J Biomol. NMR 4 (1994) 171-180.

[26] D.S. Wishart, B.D. Sykes, Chemical shifts as a tool for structure determination, Methods Enzymol. 239 (1994) 363-392.

[27] D. Kumar, A. Gautam, R.V. Hosur, A unified NMR strategy for high-throughput determination of backbone fold of small proteins, J Struct Funct Genomics 13 (2012) 201-212.

[28] D.S. Wishart, D. Arndt, M. Berjanskii, P. Tang, J. Zhou, G. Lin, CS23D: a web server for rapid protein structure generation using NMR chemical shifts and sequence data, Nucleic Acids Res. 36 (2008) W496-W502.

[29] Y. Shen and others, Consistent blind protein structure generation from NMR chemical shift data, Proc. Natl. Acad. Sci. U. S. A 105 (2008) 4685-4690.

[30] S. Raman, O.F. Lange, P. Rossi, M. Tyka, X. Wang, J. Aramini, G. Liu, T.A. Ramelot, A. Eletsky, T. Szyperski, M.A. Kennedy, J. Prestegard, G.T. Montelione, D. Baker, NMR structure determination for larger proteins using backbone-only data, Science 327 (2010) 1014-1018.

[31] L. Emsley, G. Bodenhausen, Gaussian Pulse Cascades: New Analytical Functions for Rectangular Selective Inversion and In-Phase Excitation in NMR, Chemical Physics Letters 165 (1990) 469-476.

[32] A.J. Shaka, J. Keeler, T. Frankiel, R. Freeman, An Improved Sequence for Broadband Decoupling: WALTZ-16, Journal of Magnetic Resonance 52 (1983) 335-338.

[33] A.J. Shaka, J. Keeler, R. Freeman, Evaluation of a New Broadband Decoupling Sequence: WALTZ-16, Journal of Magnetic Resonance 53 (1983) 313-340.

[34] A.J. Shaka, P.B. Barker, R. Freeman, Computer-Optimized Decoupling Scheme for Wideband Applications and Low-Level Operation, Journal of Magnetic Resonance 64 (1985) 547-552.

[35] D. Marion, M. Ikura, R. Tschudin, A. Bax, Rapid Recording of 2D NMR Spectra without Phase Cycling. Application to the Study of Hydrogen Exchange in Proteins, J. Magn Reson. 85 (1989) 393-399.

[36] M. Piotto, V. Saudek, V. Sklenar, Gradient-tailored excitation for single-quantum NMR spectroscopy of aqueous solutions, J. Biomol. NMR 2 (1992) 661-665.

[37] R. Keller, The Computer Aided Resonance Assignment Tutorial CANTINA, Verlag, Goldau, Switzerland., (2004)

[38] S.C. Panchal, N.S. Bhavesh, R.V. Hosur, Improved 3D triple resonance experiments, HNN and HN(C)N, for HN and 15N sequential correlations in (13C, 15N) labeled proteins: application to unfolded proteins, J. Biomol. NMR 20 (2001) 135-147.

[39] D. Kumar, J. Chugh, R.V. Hosur, Generation of Serine/Threonine Check Points in HN(C)N Spectra, J. Chem. Sci. 121 (2009) 955-964.

[40] A. Chatterjee, A. Kumar, R.V. Hosur, Alanine check points in HNN and HN(C)N spectra, J. Magn Reson. 181 (2006) 21-28.

[41] A. Chatterjee, N.S. Bhavesh, S.C. Panchal, R.V. Hosur, A novel protocol based on HN(C)N for rapid resonance assignment in ((15)N, (13)C) labeled proteins: implications to structural genomics, Biochem. Biophys. Res. Commun. 293 (2002) 427-432.

[42] S.A. De, A. Cavalli, S.T. Hsu, W. Vranken, M. Vendruscolo, Accurate random coil chemical shifts from an analysis of loop regions in native states of proteins, J Am. Chem Soc 131 (2009) 16332-16333.

[43] N.J. Skelton, J. Kordel, W.J. Chazin, Determination of the solution structure of Apo calbindin D9k by NMR spectroscopy, J Mol. Biol. 249 (1995) 441-462.

[44] J. Chugh, R.V. Hosur, Spectroscopic labeling of A, S/T in the 1H-15N HSQC spectrum of uniformly (15N-13C) labeled proteins, J. Magn Reson. 194 (2008) 289-294.

[45] D. Kumar, S. Paul, R.V. Hosur, BEST-HNN and 2D (HN)NH experiments for rapid backbone assignment in proteins, Journal of Magnetic Resonance 204 (2010) 111-117.

[46] D. Kumar, A. Borkar, R.V. Hosur, Facile Backbone (1H, 15N, 13Ca and 13C') Assignment of 13C/15N labeled proteins using orthogonal projection planes of HNN and HN(C)N experiments and its Automation, Magn Reson. Chem. 50 (2012) 357-363.

[47] G.T. Montelione, D. Zheng, Y.J. Huang, K.C. Gunsalus, T. Szyperski, Protein NMR spectroscopy in structural genomics, Nat. Struct Biol. 7 Suppl (2000) 982-985.

[48] M. Swain, H.S. Atreya, CSSI-PRO: a method for secondary structure type editing, assignment and estimation in proteins using linear combination of backbone chemical shifts, J Biomol. NMR 44 (2009) 185-194.

[49] S.A. De, A. Cavalli, S.T. Hsu, W. Vranken, M. Vendruscolo, Accurate random coil chemical shifts from an analysis of loop regions in native states of proteins, J. Am. Chem Soc 131 (2009) 16332-16333.

[50] S.B. Shuker, P.J. Hajduk, R.P. Meadows, S.W. Fesik, Discovering high-affinity ligands for proteins: SAR by NMR, Science 274 (1996) 1531-1534.

[51] P. Perttu, A. Arto, Sequential resonance assignment from two-dimensional inter- and intra-residue 15NΓÇô1H correlation spectra, Magnetic Resonance in Chemistry 39 (2001) 179-181.

[52] K. Pervushin, R. Riek, G. Wider, K. Wuthrich, Attenuated T2 relaxation by mutual cancellation of dipole-dipole coupling and chemical shift anisotropy indicates an avenue to NMR structures of very large biological macromolecules in solution, Proc. Natl. Acad. Sci. U. S. A 94 (1997) 12366-12371.

[53] P. Schanda, M.H. Van, B. Brutscher, Speeding up three-dimensional protein NMR experiments to a few minutes, J. Am. Chem. Soc. 128 (2006) 9042-9043.

[54] K. Pervushin, B. Vogeli, A. Eletsky, Longitudinal (1)H relaxation optimization in TROSY NMR spectroscopy, J Am. Chem Soc 124 (2002) 12898-12902.

[55] D. Rovnyak, D.P. Frueh, M. Sastry, Z.Y. Sun, A.S. Stern, J.C. Hoch, G. Wagner, Accelerated acquisition of high resolution triple-resonance spectra using non-uniform sampling and maximum entropy reconstruction, J Magn Reson. 170 (2004) 15-21.